\documentstyle[aps,prd,eqsecnum,amsfonts,amssymb,epsf,preprint,tighten]{revtex}
\draft
\title{Binary inspiral, gravitational radiation, and cosmology}
\author{Lee Samuel Finn}
\address{Department of Physics and Astronomy,\\
Northwestern University, Evanston, Illinois 60208-2900}
\date{5 July 1995}
\begin{document}
\maketitle
\widetext
\begin{abstract}
Observations of binary inspiral in a single interferometric
gravitational wave detector can be cataloged according to
signal-to-noise ratio $\rho$ and chirp mass $\cal M$. The distribution
of events in a catalog composed of observations with $\rho$ greater
than a threshold $\rho_0$ depends on the Hubble expansion,
deceleration parameter, and cosmological constant, as well as the
distribution of component masses in binary systems and evolutionary
effects. In this paper I find general expressions, valid in any
homogeneous and isotropic cosmological model, for the distribution
with $\rho$ and $\cal M$ of cataloged events; I also evaluate these
distributions explicitly for relevant matter-dominated
Friedmann-Robertson-Walker models and simple models of the neutron
star mass distribution. In matter dominated Friedmann-Robertson-Walker
cosmological models advanced LIGO detectors will observe binary neutron
star inspiral events with $\rho>8$ from distances not
exceeding approximately $2\,\text{Gpc}$, corresponding to redshifts
of $0.48$ (0.26) for $h=0.8$ ($0.5$), at an estimated rate of 1 per
week. As the binary system mass increases so does the distance it can
be seen, up to a limit: in a matter dominated Einstein-deSitter
cosmological model with $h=0.8$ ($0.5$) that limit is approximately
$z=2.7$ (1.7) for binaries consisting of two $10\,\text{M}_\odot$
black holes. Cosmological tests based on catalogs of the kind
discussed here depend on the distribution of cataloged events with
$\rho$ and $\cal M$. The distributions found here will play a pivotal
role in testing cosmological models against our own universe and in
constructing templates for the detection of cosmological inspiraling
binary neutron stars and black holes.
\end{abstract}
\narrowtext
\section{Introduction}

\subsection{Overview}

The most promising anticipated source for the United States Laser
Interferometer Gravitational-wave Observatory (LIGO)
\cite{abramovici92a}, or its French/Italian counterpart VIRGO
\cite{bradaschia90a}, is the radiation emitted during the final
moments of inspiral before the coalescence of a neutron star - neutron
star (ns-ns) binary system \cite{thorne87a}. The instruments operating
in both the LIGO and VIRGO facilities will evolve over time,
eventually becoming sensitive to neutron star binary inspirals at
distances approaching 2~Gpc~\cite{finn93a}.

Binary inspiral observations in the LIGO or VIRGO detectors will be
characterized by their signal strength and ``chirp mass'' (a
combination of the binary's component masses and cosmological
redshift). The distribution of observed inspirals with signal strength
and chirp mass depends on cosmological parameters that describe our
universe (Hubble constant, deceleration parameter, density parameter),
the distribution of neutron star masses in binary systems, the overall
density of coalescing binaries, and the properties of the detector. In
this paper I explore the binary inspiral event distribution (with
signal strength and chirp mass) in the LIGO and VIRGO detectors for
different cosmological models.

In addition to their value as quantitative expectations of what LIGO
and VIRGO can expect to observe, these {\em a priori} distributions
will play a pivotal role both in the construction of templates for
detecting binary inspirals and in the interpretation of the
observations. The distributions presented here, evaluated for our
preconceived notion of the binary inspiral rate and cosmological
parameters, are the {\em prior probabilities} required to form the
likelihood function from the observed detector output
\cite{finn93a}. Additionally, the observed inspirals will be a sample
drawn from a particular cosmological model characterized by $H_0$,
$q_0$, $\Omega_0$, neutron star mass distribution, and evolution
characteristic of our own universe. By comparing the observed
distributions to the ones described here we can measure those
properties of our own universe\footnote{A detailed study of how
accurately those measurements can be made as a function of the number
of inspiral observations is underway and will be published separately;
here I focus on describing the properties of a catalog of observations
defined by a data cut on the signal-to-noise ratio and how those
properties can be used to measure cosmological parameters.}  

These cosmological tests are analogous to the number-count tests of
classical cosmology, which, in their simplest form, involve observing
the distribution of a source population as a function of apparent
luminosity or redshift. The first suggestion that binary inspiral
number counts be used to measure interesting cosmological parameters
was made by Finn and Chernoff \cite{finn93a,chernoff93a} (although
they did not use the language usually associated with this technique
of classical astronomy). Using Monte Carlo simulations they
demonstrated that the distribution of inspiral events with signal
strength and chirp mass could be used to measure the Hubble
constant. In this work I provide a more general, explicit, and
complete exposition of the properties of binary inspiral observation
catalogs.

The cosmological implications of gravitational wave observations of
binary inspiral were first recognized by Schutz \cite{schutz86a}. He
pointed out that each inspiraling binary is a standard candle in the
sense that, if observed in three independent interferometers, its
luminosity distance can be determined from the observed detector
response.  If an observed inspiral is associated with one of the
several galaxy clusters that reside in its positional error box (whose
determination also requires three interferometers), and if the
redshifts of those clusters are determined optically, then observation
of several inspiraling binaries would lead to a statistical
determination of the Hubble constant that is independent of the cosmic
distance ladder and the uncertainties that lurk therein.

Markovi\'c \cite{markovic93a} proposed a variation on the general
theme introduced by Schutz: he observed that known neutron star masses
were all close to $1.4\,\text{M}_\odot$ and that, in any event, there
is a {\em maximum} neutron star mass. The observed chirp mass is a
function of the mass of the binary's two components and its redshift.
Assuming that the mass distribution in neutron star binaries does not
evolve significantly over the range of binary inspiral observations,
examination of the chirp mass distribution in binary systems at fixed
luminosity distance would reveal the corresponding redshift. Thus,
gravitational radiation observations alone might suffice to determine
the Hubble constant.

Unfortunately, detailed calculations show that, even for the most
advanced LIGO and VIRGO detectors that have been discussed, the
fractional uncertainty in the measured luminosity distance will be of
order unity for events seen more frequently than thrice per year ({\em
i.e.,} for events at distances greater than approximately
$100\,\text{Mpc}$) \cite{jaranowski92a,cutler94a}, and the angular
position error boxes for these events are likewise large (on order
$10\,\deg^2$ \cite{wipf95a}). Consequently, cosmological tests that
rely on accurate and precise measurements of the distance and position
of inspiraling binaries using LIGO and VIRGO are not promising. 

In contrast, the cosmological tests discussed here and in
\cite{chernoff93a} require only gravitational wave observation in a
single interferometer. Furthermore, advanced LIGO detectors can expect
to observe approximately 50 ns-ns binary inspiral events per year,
from distances up to $2\,\text{Gpc}$, whose signal strength can be
measured to better than 10\% and whose chirp mass can be measured to
better than $0.1\%$ \cite{finn93a,cutler94a}. The rate, depth,
accuracy and precision of these single interferometer observations
suggest that cosmological tests based on the distribution of observed
events with signal strength and chirp mass have great promise.

\subsection{Outline}

In this paper I calculate the expected properties of a catalog of
binary inspiral observations made by a single interferometric
gravitational wave detector. A catalog is presumed to contain a record
of all binaries that coalesced during the observation period and whose
inspiral signal-to-noise ratio $\rho$ was greater than the catalog
limit $\rho_0$. The catalog properties depend on the coalescence rate
density and the binary system component mass distribution, both of
which may vary with redshift. Notation for these and a discussion of
the observations that bear on them is the topic of section
\ref{sec:dN/dtdMdV}.

In section \ref{sec:snr} I discuss the signal-to-noise ratio $\rho$
as a measure of signal strength. The signal-to-noise ratio of a
particular inspiraling binary depends on the binary's intrinsic
properties, its distance from and orientation with respect to the
detector, and the detector's intrinsic properties. The way in which
$\rho$ depends on the signal and detector properties suggests a useful
measure of the detector bandwidth, which I discuss here as
well. Finally, section \ref{sec:snr} concludes by specializing the
discussion to the specific properties of the proposed LIGO and VIRGO
gravitational radiation detectors.

The important properties of a binary inspiral observation catalog are
the distribution of cataloged events with $\rho$ and the binary system
``chirp mass'' $\cal M$. These distributions depend on the
cosmological model, which includes the evolution with redshift of the
neutron star mass distribution and the coalescing binary number
density. In section \ref{sec:discussion} I give general expression for
these distributions and discuss how they may be used together with
actual observations to test cosmological models. Also in this section
I give expressions for the {\em catalog depth} (the maximum redshift
of a binary system that can have $\rho$ greater than $\rho_0$) and the
total rate that proposed interferometers can expect to observe
inspiraling binaries with $\rho$ greater than $\rho_0$. All of these
general expressions are evaluated explicitly for relevant
matter-dominated Friedmann-Robertson-Walker cosmological models and a
simple model of the neutron star mass distribution. Finally, I
summarize my conclusions in section \ref{sec:conclusions}.

\section{Coalescence rate density}
\label{sec:dN/dtdMdV}

\subsection{Introduction}

The signal-to-noise ratio $\rho$ of a binary inspiral in a LIGO-like
interferometer depends on the relationship between the binary and the
detector ({\em i.e.,} orientation, distance and redshift) and also on
certain intrinsic properties of the system ({\em i.e.,} component
masses and spins). Of these intrinsic properties, the {\em intrinsic
chirp mass,}
\begin{equation}
{\cal M}_0 \equiv \mu^{3/5} M^{2/5}
\end{equation}
where $\mu$ and $M$ are the binary's reduced and total mass, plays the
most important role: all the other intrinsic properties offer only
small corrections to $\rho$.

Gravitational-wave detectors like LIGO or VIRGO do not measure ${\cal
M}_0$; instead, they measure
\begin{equation}
{\cal M} \equiv {\cal M}_0 (1+z) \label{eqn:M},
\end{equation}
where $z$ is the system's redshift with respect to the detector. To
distinguish between $\cal M$, which involves the system's redshift,
and ${\cal M}_0$, which depends only on the binary's intrinsic
properties, I refer to the former as the {\em observed} chirp mass, or
simply the chirp mass, and the latter as the {\em intrinsic} chirp
mass.

In order to describe the binaries included in a signal-to-noise
limited catalog we must first describe the coalescing binary
distribution in space and in ${\cal M}_0$. The notation I use to
describe this distribution is defined in section \ref{sec:calN}, while
in section \ref{sec:NConstraints} I discuss what is known about the
distribution from present-day observations.

\subsection{Definitions and notation}\label{sec:calN}

Assume that coalescing binaries are distributed homogeneously and
isotropically with the cosmological fluid and define the binary
coalescence {\em local specific rate density} $\cal N$ by
\begin{equation}
{\cal N} \equiv {d^3N\over dt\, dV\, d{\cal M}_0},
\end{equation}
where $dV$ is a {\em co-moving} cosmological fluid volume element and
$dt$ is a proper time interval measured in the fluid rest frame.  The
total co-moving rate density on the surface of homogeneity at redshift
$z$ is thus
\begin{equation}
\dot{n}(z) = \int {\cal N}({\cal M}_0,z)\, d{\cal M}_0.
\end{equation}

Define the ratio of the total co-moving rate density at a redshift $z$
[$\dot{n}(z)$]
to that at the present epoch [$\dot{n}_0=\dot{n}(z=0)$] by ${\cal E}$:
\begin{equation}
\dot{n}(z) = {\cal E}(z)\dot{n}_0.
\end{equation}
The distribution of coalescing binaries with intrinsic chirp mass
${\cal M}_0$ on the surface at redshift $z$ is
\begin{equation}
{\cal P}({\cal M}_0|z) = {{\cal N}({\cal M}_0,z)\over\dot{n}(z)}, 
\end{equation}
where, by construction, 
\begin{equation}
1 = \int d{\cal M}_0\, {\cal P}({\cal M}_0|z).
\end{equation}
The homogeneous and isotropic local specific rate density can thus be
written
\begin{equation}
{\cal N}({\cal M}_0,z) = \dot{n}_0{\cal E}(z){\cal P}({\cal M}_0|z).
\end{equation}

Since we have defined $\cal N$ and $\dot{n}$ in terms of the co-moving
volume element and intrinsic chirp mass, in the absence of evolution
${\cal E}$ is unity and ${\cal P}$ is independent of
$z$. Additionally, for infinitesimal $d{\cal M}_0$, ${\cal P}({\cal
M}_0|z)d{\cal M}_0$ can be interpreted as the probability that a
randomly chosen binary on the surface at redshift $z$ has intrinsic
chirp mass in the range ${\cal M}_0$ to ${\cal M}_0+d{\cal M}_0$.

\subsection{Observational constraints on $\cal N$}
\label{sec:NConstraints}

\subsubsection{Rate density at the current epoch: $\dot{n}_0$}
\label{sec:n0}

Cosmological tests that depend on the observed {\em distribution} of
inspirals with $\rho$ and/or $\cal M$ do not depend on
$\dot{n}_0$. Nevertheless, it is necessary to know $\dot{n}_0$ in
order to estimate how long it will take to accumulate a catalog of
observations large enough that such tests will give meaningful
results.

The best current estimate of the ns-ns binary coalescence rate density
at the current epoch is
$1.1\times10^{-8}h\,\text{Mpc}^{-3}\;\text{yr}^{-1}$, where $h$ is the
Hubble constant measured in units of
$100\,\text{Km}\,\text{s}^{-1}\,\text{Mpc}^{-1}$
\cite{narayan91a,phinney91a}. This estimates relies on the 3 observed
binary pulsar systems that will coalesce in less than a Hubble time
(PSRs~1913+16, 1534+12, and 2127+11C
\cite{taylor89a,wolszczan91a,prince91a}). Since the number of observed
systems is small the actual rate is quite uncertain: Phinney
\cite{phinney91a} has estimated that, while unlikely, a rate two
orders of magnitude higher or lower could be reconciled with current
observations.

Black hole - black hole (bh-bh) and bh-ns binaries are believed to
form at rates comparable to ns-ns binaries; however, the masses of
these systems and the fraction that merge in less than a Hubble time
are entirely uncertain \cite{narayan91a,phinney91a}.

\subsubsection{Intrinsic chirp mass distribution and evolution: $\cal
P$ and $\cal E$} \label{sec:P&E}

The intrinsic chirp mass distribution ${\cal P}({\cal M}_0|z)$ depends
on the binary system component mass distribution on the slice of
homogeneity at redshift $z$. Denote a binary's component masses 
as $m_1$ and $m_2$ and write their joint probability density on a
surface of redshift $z$ as $P(m_1,m_2|z)$. The intrinsic chirp mass
distribution ${\cal P}({\cal M}_0|z)$ on that slice is then 
\begin{eqnarray}
{\cal P}({\cal M}_0|z) &=& \int\!\!\!\int 
dm_1\,dm_2\,
P(m_1,m_2|z)\nonumber\\
&&\qquad{}\times\delta\!\left(\mu^{3/5}M^{2/5}-{\cal M}_0\right).
\end{eqnarray}
The determination of ${\cal P}({\cal M}_0|z)$ thus reduces to finding
$P(m_1,m_2|z)$. 

Both theoretical and observational evidence suggest that the neutron
star mass distribution is narrow \cite{woosley92a,finn94b,bethe95a}. A
simple model of the mass distribution has the component masses in a
binary uncorrelated and uniformly distributed between upper and lower
bounds $m_u$ and $m_l$; then
\begin{eqnarray}
P(m_1,m_2|m_l,m_u) &=& P(m_1|m_l,m_u)P(m_2|m_l,m_u) \nonumber\\
&=& \left(m_u-m_l\right)^{-2}
\end{eqnarray}
and 
\begin{eqnarray}
\lefteqn{P({\cal M}_0|m_l, m_u) = }&&\nonumber\\
&&\qquad{}\int_{m_l}^{m_u}\!\int_{m_l}^{m_u} dm_1\,dm_2
{\delta\left[\left(m^3_2 m^3_1\over m_1+m_2\right)^{1/5}-{\cal
M}_0\right]
\over
\left(m_u-m_l\right)^2
}. \label{eqn:p(m0)}
\end{eqnarray}
where $m_l$ and $m_u$ may depend on $z$. The probability density
$P({\cal M}_0|m_l,m_u)$ is maximized when ${\cal M}_0=(m_l
m_u)^{3/5}/(m_l+m_u)^{1/5}$.

The limited observations of the neutron star mass distribution in
binary pulsar systems provide independent 95\% confidence intervals
for $m_l$ and $m_u$ \cite{finn94b}:
\begin{equation}
\begin{array}{rcccl}
1.01&<&m_l/\text{M}_\odot&<&1.34\\
1.43&<&m_u/\text{M}_\odot&<&1.64.
\end{array}
\end{equation}
The most likely values of $m_l$ and $m_u$ are
\begin{equation}
\begin{array}{rcl}
m_l&=&1.29\,\text{M}_\odot,\\
m_u&=&1.45\,\text{M}_\odot.
\end{array}
\end{equation}
Over this narrow range $P({\cal M}_0|m_l,m_u)$ is, to an excellent
approximation, piecewise linear:
\begin{eqnarray}
\lefteqn{P({\cal M}_0|m_l, m_u) \simeq}&&\nonumber\\
&&\qquad{2\over m_>-m_<}
\left\{
\begin{array}{ll}
{{\cal M}_0-m_<\over m_0-m_<}&\text{if $m_0>{\cal M}_0>m_<$,}\\
{m_>-{\cal M}_0\over m_>-m_0}&\text{if $m_>>{\cal M}_0>m_0$,}\\
0&\text{otherwise}
\end{array}
\right.\label{eqn:approx-p(m0)}
\end{eqnarray}
where
\begin{mathletters}
\begin{eqnarray}
m_< &\equiv& m_l/2^{1/5},\\
m_0 &\equiv& {\left(m_l
m_u\right)^{3/5}\over\left(m_l+m_u\right)^{1/5}}\\
m_> &\equiv& m_u/2^{1/5}.
\end{eqnarray}
\end{mathletters}

At present we can observe only nearby binary pulsar systems;
consequently, there are no observations that bear directly on the
variation of $\cal E$ or $\cal P$ with $z$. Theoretical studies
suggest that the initial mass of neutron stars formed by stellar core
collapse do not vary significantly with the progenitor mass or
composition \cite{woosley92a}. After formation the mass may evolve
owing to accretion from a companion; however, in any event it is not
likely that either ${\cal P}$ or ${\cal E}$ vary with $z$ more rapidly
than do galaxies. In section \ref{sec:discussion} I provide general
expressions for and detailed examples of the expected distribution of
events in a catalog of binary inspiral events; for the detailed
examples I neglect evolution in $\cal E$ and $\cal P$ entirely. As
shown in section \ref{sec:depth}, advanced LIGO and VIRGO detectors
will observe ns-ns binaries from redshifts not expected to exceed
$z\simeq0.5$ and with the preponderance of events arising from
$z\simeq0.1$; consequently, neglect of evolution is not an
unreasonable approximation.

\section{The signal-to-noise ratio}\label{sec:snr}

\subsection{Outline}

The signal-to-noise ratio measures the signal amplitude in terms of a
detector's noise properties. In subsection \ref{sec:snr/subsec:intro}
I define the signal-to-noise ratio $\rho$ and discuss the subtle issue
of how $\rho$ is estimated, but not determined, by observation. The
functional form of $\rho$ depends on the detector response to binary
inspiral radiation, which I describe in subsection
\ref{sec:response}. In subsection \ref{subsec:snr} I give the binary
inspiral signal-to-noise ratio in terms of the same parameters that
characterize the detector response. The form of $\rho$ suggests a
natural definition of a detector's effective bandwidth for binary
inspiral observations; I discuss this bandwidth function in subsection
\ref{subsec:snr} as well.

The relative orientation of the detector and binary is described by a
function of four angles. While these angles cannot be measured by
observations in a single interferometer, important properties of the
angular orientation function can be calculated independently of the
particulars of the binary or the detector. These properties play an
important role in determining the binary inspiral catalog properties
and interpreting individual observations. I discuss the angular
orientation function in subsection \ref{sec:pTheta}.

Finally, in subsection \ref{sec:ligo/virgo} I specialize the
discussion of the signal-to-noise to the proposed LIGO and VIRGO
gravitational radiation detectors.

\subsection{The signal-to-noise ratio: general
comments}\label{sec:snr/subsec:intro} 

Let $s(t)$ be the detector response to a gravitational radiation
signal from any source. If the detector noise is Gaussian with
one-sided power spectral density $S_n(f)$ then the signal-to-noise
ratio $\rho^2$ is {\em defined} to be
\begin{equation}
\rho^2 = 2\int_{-\infty}^{\infty} {\widetilde{s}(f)\widetilde{s}^*(f)\over
S_n(|f|)} df,\label{eqn:snr}
\end{equation}
where $\widetilde{s}$ is the Fourier transform of the detector response,
\begin{equation}
\widetilde{s}(f) \equiv \int_{-\infty}^\infty df\,e^{2\pi i f t} s(t),
\end{equation}
and $\widetilde{s}^*$ is its complex conjugate \cite{finn92a}.

An observation of a gravitational wave signal in a noisy detector
entails a measurement of the signal properties in the presence of a
particular instance of the detector noise. Analysis of the detector
output results in an {\em estimate} of $\rho$ as well as other
parameters that describe $s(t)$. {\em Throughout this work I use
$\rho$ to mean the actual signal-to-noise ratio, as defined by
equation \ref{eqn:snr}, and not the estimate that arises in an
observation.}

In section \ref{sec:discussion} I find the distribution of sources
with $\rho$ and $\cal M$ in different cosmological models. By
comparing these distributions with the observed one we can determine
the model that best describes our own universe. In making that
comparison it is critical to distinguish between $\rho$ and $\cal M$
(as {\em defined} by equations \ref{eqn:snr} and \ref{eqn:M}) and the
{\em estimates} of $\rho$ and $\cal M$ that results from observations
made in a detector.

{\em The estimate that results from an observation is a probability
distribution $P$ for the parameters that describe the signal} --- in
the case of binary inspiral in a single interferometer, these include
$\rho$ and the chirp mass $\cal M$ (see sec.~\ref{sec:response}
below). The probability distribution associated with an observation is
generally not reported; instead, what is reported most often is a set
of {\em estimators} that characterize the distribution and its
moments. Among the most popular is the the maximum likelihood
estimator, which is the set of parameter values that maximize $P$.

Estimators are summaries of the distribution $P$ and their utility
depends on how accurately they are able to represent it. When $P$ is
very sharply peaked ({\em i.e.,} there is little uncertainty in the
measurement) then it may be approximated by a $\delta$-function and
summarized accurately by the maximum likelihood
estimator. When $P$ is sharply peaked but with a not insignificant
width, then it may be approximated near its peak by a Gaussian and
accurately represented by the maximum likelihood estimators and their
covariance. When the distribution is not sharply peaked, however, then
the more general uncertainties reflected in the detailed structure of
$P$ play an essential role in the observation's interpretation and no
summary of $P$ is especially useful.

That a small set of estimators built from $P$ for a particular
observation do not provide a useful summary does not mean that the
observation itself is unreliable or uninformative; rather, it means
only that greater care must be taken in its interpretation.  Finn
\cite[sec. (c)]{finn94a} gives an example of how the the probability
distribution resulting from an observation should be used in the
interpretation of neutron star mass observations in binary pulsar
systems; a further discussion of this point in the context of
cosmological tests using binary inspiral observation catalogs is part
of a work in preparation.

\subsection{Detector response to binary inspiral}\label{sec:response}

The detector response $s(t)$ to the gravitational radiation from an
inspiraling binary system depends on the distance and relative
orientation of the source and the detector, as well as on certain
intrinsic properties of the binary. The relative orientation of the
source and the detector is described by four angles: two ($\theta$ and
$\phi$) describe the direction to the binary relative to the detector,
and two ($i$ and $\psi$) describe the binary's orientation relative to
the line-of-sight between it and the detector.

To describe $\theta$ and $\phi$, consider a single interferometric
gravitational wave detector whose arms form a right angle. Let the
arms themselves determine the $\bbox{x}$ and $\bbox{y}$ axes of a
right-handed Cartesian coordinate system with the $\bbox{z}$-axis
pointing skyward. In this coordinate system the gravitational waves
from an inspiraling binary arrive from the direction $\bbox{n}$, which
can be defined in terms of the spherical coordinates $\theta$ and
$\phi$ in the usual way:
\begin{mathletters}
\begin{eqnarray}
\cos\theta &\equiv& -\bbox{n}\cdot\bbox{z},\\
\tan\phi &\equiv& {\bbox{n}\cdot\bbox{y}\over\bbox{n}\cdot\bbox{x}}.
\end{eqnarray}

To describe $i$ and $\psi$, let $\bbox{J}$ represent the total angular
momentum of the binary system. The detector response $s(t)$ depends on
the {\em inclination angle} $i$ between $\bbox{J}$ and $\bbox{n}$,
\begin{equation}
\cos i \equiv -\bbox{J}\cdot\bbox{n}/|\bbox{J}|,
\end{equation}
and the orientation $\psi$ of the angular momentum about
$\bbox{n}$,
\begin{equation}
\cot\psi \equiv
{
\bbox{J}\cdot\bbox{n}\times\bbox{z}
\over
\bbox{J}\cdot\left[\bbox{z}-\bbox{n}(\bbox{z}\cdot\bbox{n})\right]
}.
\end{equation}
\end{mathletters}
The conventions for the orientation angles $\theta$, $\phi$, $i$, and
$\psi$ described here are the same as those used in
\cite{thorne87a,finn93a} (note, however, that the {\em description} of
the angles in \cite{finn93a} is incorrect).

At Newtonian ({\em i.e.,} quadrupole formula) order the only intrinsic
property of the binary system that affects the waveform is ${\cal
M}_0$. At this order the detector response (a dimensionless strain) to
the binary inspiral signal is
\begin{mathletters}
\begin{equation}
s(t) = \left\{
\begin{array}{ll}
{{\cal M}\over d_L}\Theta \left(\pi f{\cal
M}\right)^{2/3}\cos\left[\chi + \Phi(t)\right] &\text{for $t<T'$},\\
0&\text{for $t>T'$},
\end{array}
\right.
\label{eqn:response}
\end{equation}
where $\chi$ is a constant phase, 
\begin{eqnarray}
\Theta &\equiv& 2\left[
F_{{}+{}}^2\left(1+\cos^2i\right)^2
+ 4F_{\times}^2\cos^2i\right]^{1/2},\\
F_{{}+{}} &\equiv& 
{1\over2}\left(1+\cos^2\theta\right)\cos2\phi\cos2\psi \nonumber\\
&&\qquad{}- \cos\theta\sin2\phi\sin2\psi, \\
F_\times &\equiv&
{1\over2}\left(1+\cos^2\theta\right)\cos2\phi\sin2\psi \nonumber\\
&&\qquad{}+ \cos\theta\sin2\phi\cos2\psi, \\
f &=& {1\over\pi\cal M}\left[{5\over 256}{{\cal M}\over
T-t}\right]^{3/8},\\
\Phi &\equiv& 2\pi\int_T^t f(t) dt = -2\left({T-t\over5{\cal
M}}\right)^{5/8}, 
\end{eqnarray}
\end{mathletters}
$d_L$ is the source's luminosity distance, $T'$ is the moment when the
inspiral waveform terminates (either because the binary components
have coalesced or because the orbital evolution is no longer
adiabatic), and $T>T'$ would be the moment of coalescence if the two
components of the binary system were treated as point masses in the
quadrupole approximation (the difference between $T$ and $T'$ is small
but not negligible). Note that dependence of the response on the
angles $\theta$, $\phi$, $i$, and $\psi$ is contained entirely in the
orientation function $\Theta$, which does not depend on any other
properties of the binary or the interferometric detector. This
important point will be discussed further in subsection
\ref{sec:pTheta}.

Post-Newtonian corrections do not contribute significantly to the
signal-to-noise ratio for solar-mass binary inspiral in the LIGO and
VIRGO interferometers. For symmetric binaries ({\em i.e.,} those with
equal or near equal mass components) the first post-Newtonian
correction is proportional to $M/r$, where $M$ is the binary's total
mass and $r$ the component separation. Advanced LIGO and VIRGO
interferometers are expected to be most sensitive to binary inspiral
radiation in the bandwidth 20--$200\,\text{Hz}$ (see
sec.~\ref{sec:bandwidth}) and, in this bandwidth, $M/r\lesssim4\%$ for
solar mass binaries. Post-Newtonian effects are more important for
more massive binaries, since at fixed quadrupole radiation frequency
$M/r$ is greater for greater $M$:
\begin{mathletters}
\begin{eqnarray}
\left({M\over r}\right) &\simeq& \left(\pi f M\right)^{1/3}\\
&\simeq&
0.042\,\left({f\over200\,\text{Hz}}{M\over2.8\,\text{M}_\odot}\right)^{2/3}\\
&\simeq&
0.16\,\left({f\over200\,\text{Hz}}{M\over20\,\text{M}_\odot}\right)^{2/3}.
\end{eqnarray}
\end{mathletters}
Post-Newtonian effects are more important for binaries with extreme
mass ratios: in general, the first post-Newtonian correction is
proportional to $(\delta m/M)(M/r)^{1/2}$, where $\delta m$ is the
component mass difference. Finally, spin-orbit coupling in systems
whose components have large spin angular momenta can lead to orbital
precession and a waveform modulation which can affect $\rho$
significantly \cite{apostolatos94a}.

That the quadrupole waveform can be used to {\em estimate} $\rho$ for
an inspiraling binary does not mean it can also be used to {\em
detect} a binary inspiral signal in the detector output. Detection, or
estimation of signal parameters from observation, involves the
comparison of the detector output with a model of the detector
response to the radiation. ``Small'' differences between the actual
and modeled signal evolution, particularly its phase, can have a large
effect on the measured value of the signal parameters, including the
signal-to-noise ratio estimated from the observations
\cite{finn93a,cutler93b}.

\subsection{Binary inspiral signal-to-noise ratio}\label{subsec:snr}

The signal-to-noise ratio $\rho$ (eq. \ref{eqn:snr}) corresponding to
the detector response $s(t)$ (eq. \ref{eqn:response}) is
\cite{finn93a}
\begin{mathletters}
\begin{eqnarray}
\rho &=& 8 \Theta {r_0\over d_L} 
\left({\cal M}\over1.2\,\text{M}_\odot\right)^{5/6}
\zeta(f_{\text{max}}), \label{eqn:rho}
\end{eqnarray}
where 
\begin{eqnarray}
r_0^2 &\equiv& {5\over192\pi}\left(3\over20\right)^{5/3}
x_{7/3} \text{M}^2_\odot , \label{eqn:r0} \\
x_{7/3} &\equiv& \int_0^\infty
{df\,\left(\pi\text{M}_\odot\right)^2\over
\left(\pi f \text{M}_\odot\right)^{7/3} S_n(f)},
\label{eqn:x73}\\
f_{\max} &=& f(T')/2,\label{eqn:fmax0}\\
\zeta(f_{\text{max}}) &\equiv& 
{1\over x_{7/3}}
\int_0^{2f_{\text{max}}}
{df\,\left(\pi\text{M}_\odot\right)^2\over
\left(\pi f \text{M}_\odot\right)^{7/3} S_n(f)}.
\label{eqn:zeta}
\end{eqnarray}
\end{mathletters}
In equation \ref{eqn:rho} $\Theta$, ${\cal M}$ and $d_L$ depend on the
particular binary system under consideration and $r_0$ is a
characteristic distance that depends only on the detector's noise
power spectral density $S_n(f)$. The dimensionless function $\zeta$,
which also depends only on the detector's noise spectrum, increases
monotonically from 0 to 1. Its argument $f_{\text{max}}$ is the
redshifted instantaneous orbital frequency when the inspiral
terminates (at $t=T'$) either because the compact components have
coalesced or because the orbital evolution is no longer adiabatic and
coalescence is imminent.

\subsubsection{$r_0$ and $\zeta$}

The characteristic distance $r_0$ and the function $\zeta$ describe
different aspects of an interferometric detectors sensitivity to
binary inspiral gravitational radiation. For a fixed binary, the
larger $r_0$, the greater $\rho$; for fixed $\rho$, the larger $r_0$,
the farther the detector can ``see.''  Decreasing the noise power in
any band increases $r_0$, but (owing to the factor $f^{-7/3}$ in the
integrand of the expression for $x_{7/3}$) improvements at low
frequency are more effective than those high frequency.

The function $\zeta$ reflects the overlap of the signal power with the
detector bandwidth.  The orbit of an inspiraling compact binary
evolves adiabatically owing to gravitational radiation emission until
an innermost circular orbit (ICO) is reached at an instantaneous
orbital frequency $f_{\text{ICO}}$. At the ICO the orbit evolves on a
dynamic timescale and the components coalesce quickly. Thus, the
adiabatic inspiral waveform ends when the orbital frequency reaches
$f_{\text{ICO}}$\footnote{The radiation waveform from the final plunge
and the early stages of coalescence is not yet known. If, after
coalescence, a black hole forms, then the final radiation reflects the
black hole's quasinormal modes, which are damped rapidly. The inspiral
waveform, and with it our ability to model the detector response, ends
when the orbital frequency reaches $f_{\text{ICO}}$.}. The quadrupole
radiation frequency observed at the detector when the orbital
frequency is $f_{\text{ICO}}$ is
\begin{equation}
2f_{\text{max}}={2f_{\text{ICO}}\over 1+z} 
\end{equation}
and the inspiral signal termination is represented in the detector
response by the cut-off at $t>T'$ (see eqs.~\ref{eqn:response} and
\ref{eqn:fmax0}).

Since the binary orbit evolves adiabatically from low frequencies to
$f_{\text{ICO}}$, the observed quadrupole radiation spectrum has
significant power only up to frequency $2f_{\text{max}}$. If
$2f_{\text{max}}$ is very much greater than the frequency where the
detector noise is minimized, than the signal power bandwidth overlaps
completely with the detector bandwidth and
$\zeta(f_{\text{max}})\simeq1$. On the other hand, if
$2f_{\text{max}}$ is much less than the frequency where the detector
noise is minimized, than the overlap of the signal power and detector
bandwidths is negligible, $\zeta(f_{\text{max}})\simeq0$ and
$\rho\simeq0$. 

The orbital frequency at the transition of the binary orbit from
adiabatic inspiral to plunge and coalescence has been studied using
high-order post-Newtonian methods \cite{kidder93a}. For symmetric
binaries ({\em i.e.,} those with equal-mass components) the
instantaneous redshifted instantaneous orbital frequency of the ICO is
given by
\begin{mathletters}
\label{eqn:fmax}
\begin{eqnarray}
f_{\text{max}} &=& {f_{\text{ICO}}\over1+z},\\
&=& {710.\,\text{Hz}\over1+z}
\left(2.8\,\text{M}_\odot\over M\right),\label{eqn:fmax-ns}\\
&=& {99.\,\text{Hz}\over1+z}
\left(20.\,\text{M}_\odot\over M\right)\label{eqn:fmax-bh},
\end{eqnarray}
\end{mathletters}
where $M$ is the binary's total mass. More generally, for binaries
with compact components $f_{\text{ICO}}$ depends inversely on $M$ and
directly on a function of the dimensionless ratio $\mu/M$; Kidder
\cite[fig.~4]{kidder93a} shows $Mf_{\text{ICO}}$ for binary systems of
arbitrary mass asymmetry. For a ns-ns binary the component's proper
separation at the ICO is greater than a neutron star diameter; so,
coalescence occurs after the transition from inspiral to plunge
\cite{kidder93a}. Tidal dissipation is important in determining the
inspiral rate only in the last few orbits before contact
\cite{kochanek92a,bildsten92a}; consequently, equation \ref{eqn:fmax}
is applicable for ns-ns binaries and should be applicable for black
hole binaries as well.

\subsubsection{Detector bandwidth and data analysis
templates}\label{sec:bandwidth}

The probability that a signal with detector response $s(t)$ is present
in the output $g(t)$ of a noisy detector is related to
\begin{mathletters}
\label{eqn:prob}
\begin{equation}
2\left<g,s\right>-\left<s,s\right>,
\end{equation}
where
\begin{equation}
\left<g,h\right> \equiv \int_{-\infty}^\infty
df{\widetilde{g}(f)\widetilde{h}^*(f)\over S_n(|f|)}.
\end{equation}
\end{mathletters}
The actual signal, and consequently the detector response $s$, may be
difficult or impossible to evaluate exactly (this is certainly the
case for binary inspiral); so, we would like to be able to use an
approximate model $S(t)$ in lieu of the actual detector response $s$
in equations \ref{eqn:prob}. From equations \ref{eqn:prob} it is clear
that any approximate model $S(t)$ may be used as long as it matches
$\widetilde{s}(f)$ closely wherever $|\widetilde{s}|^2/S_n(f)$ is relatively
large. This latter quantity is just the fraction of $\rho^2$
contributed by the signal power at frequency
$f$ and is proportional to $\zeta'$. This suggests that we define
the {\em bandwidth function}
\begin{equation}
{\cal B}(f) \equiv \left\{\begin{array}{ll} 
{f\over2}{\zeta'({f/2})\over\zeta(f_{\text{max}})}&f<2f_{\text{max}}\\
0&f>2f_{\text{max}}
\end{array}\right.\label{eqn:bwidth}
\end{equation}
which is the fraction of the signal-to-noise ratio contributed by
signal power at frequency $f$ in a logarithmic frequency bandwidth.
{\em An approximate detector response model $\widetilde{S}$ may be
used instead of the actual response $\widetilde{s}$ as long as
$\widetilde{S}$ accurately reflects $\widetilde{s}$ wherever $\cal B$
is large.} Knowing where and how $\widetilde{S}$ needs to be accurate
can simplify greatly the construction of approximate templates for
data analysis; for this purpose $\cal B$ should prove a useful guide.

\subsection{Distribution of the orientation function
$\Theta$}\label{sec:pTheta} 

The signal-to-noise ratio of an inspiraling binary in a single
interferometric detector depends on the relative orientation of the
source and the detector through the angles $\theta$, $\phi$, $i$, and
$\psi$. The dependence of $\rho$ on these angles is confined to the
angular orientation function $\Theta$, which is independent of all
other properties of the binary system.  From observations of binary
inspiral in a single interferometer we can measure $\rho$ and $\cal M$
but not $\Theta$.  Even though $\Theta$ cannot be measured, because it
depends only on $\theta$, $\phi$, $i$ and $\psi$ we actually know
quite a bit about it: since, averaged over many binaries,
$\cos\theta$, $\phi/\pi$, $\cos i$, and $\psi/\pi$ are all
uncorrelated and uniformly distributed over the range $(-1,1)$ we know
the probability that $\Theta$ takes on any value. This probability
distribution is found numerically in \cite{finn93a}; here I note only
that $0\leq\Theta\leq4$ and that to an excellent approximation
\begin{equation}
P_\Theta(\Theta) = \left\{
\begin{array}{ll}
5\Theta\left(4-\Theta\right)^3/256&
\text{if $0<\Theta<4$}\\
0&\text{otherwise.}
\end{array}\right.
\label{eqn:pTheta}
\end{equation}

To determine the binary coalescence rate $\rho$ greater than $\rho_0$
we need to know $\cal N$ and also how the signal-to-noise ratio of
binaries with intrinsic chirp mass ${\cal M}_0$ on a surface at
redshift $z$ is distributed. This latter distribution,
$P_\rho(\rho|{\cal M}_0,z)$, is related to $P_\Theta$:
\begin{eqnarray}
P_\rho(\rho|{\cal M}_0,z,{\cal C},{\cal D})&=&
P_\Theta\left[\Theta(\rho)\right]
\left.{\partial\Theta\over\partial\rho}\right|_{{\cal M}_0,z}
\nonumber\\
&=&
P_\Theta\left[
{\rho\over8}{d_L(z)\over r_0\left(1+z\right)^{5/6}}
\left(1.2\,\text{M}_\odot\over{\cal M}_0\right)^{5/6}
\right] \nonumber\\
&&\qquad{}\times
{d_L\left(1.2\,\text{M}_\odot/{\cal M}_0\right)^{5/6}\over 
8 r_0\left(1+z\right)^{5/6}},
\label{eqn:p(rho)}
\end{eqnarray}
where $\cal C$ represents the cosmological model ($H_0$, $q_0$,
$\Omega_0$, $\cal P$, $\cal E$) and $\cal D$ the detector ({\em i.e.,}
$r_0$ and $\zeta$). 

\subsection{$r_0$ and $\zeta$ for the LIGO and VIRGO
detectors}\label{sec:ligo/virgo} 

LIGO will consist of three interferometers: one in Livingston,
Louisiana and two in Hanford, Washington. The Louisiana interferometer
and one of the Washington interferometers will have 4~Km armlengths;
the second Washington interferometer will have a 2~Km armlength and
share the same vacuum system as the 4~Km interferometer. For the
proposed LIGO interferometers as described in \cite{ligo89a} and
modeled in \cite{finn93a}, $r_0$ ranges from $13\,\text{Mpc}$ (for
``initial'' interferometers) to $237\,\text{Mpc}$ (for ``advanced''
interferometers)\footnote{The noise model used in \cite{ligo89a} and
\cite{finn93a} assumed a thermal noise spectrum corresponding to
viscous damping in the pendulum suspensions and the internal modes of
the test masses. It is now recognized that these modes are structure
damped \cite{saulson90a,gillespie94a,gillespie94b} with a
correspondingly different noise spectrum. Preliminary estimates
indicate that this improved noise model reduces $r_0$ for initial LIGO
interferometers, but leaves it unchanged for advanced
ones.\label{fnote:noise}}. As the LIGO detectors develop, incremental
improvements will increase $r_0$.

The orientation of the Washington and Louisiana interferometers were
chosen to be as close to parallel as possible; consequently, a simple
approximation treats the LIGO {\em detector} (all three
interferometers operating in ``triple-coincidence'') as a single
interferometer of arm length
\begin{equation}
\left[4^2 + 4^2 + 2^2\right]^{1/2}\text{Km} = 6\,\text{Km.}
\end{equation}
For this ``super-interferometer'' $r_0$ ranges from $19\,\text{Mpc}$
to $355\,\text{Mpc}$. 

Here and below, reference to the LIGO {\em detector} refers to the
three interferometers operating as a single detector, while reference
to a LIGO {\em interferometer} indicates one of the 4~Km
interferometers operating in isolation.

\begin{figure}
\epsfxsize=\columnwidth\epsffile{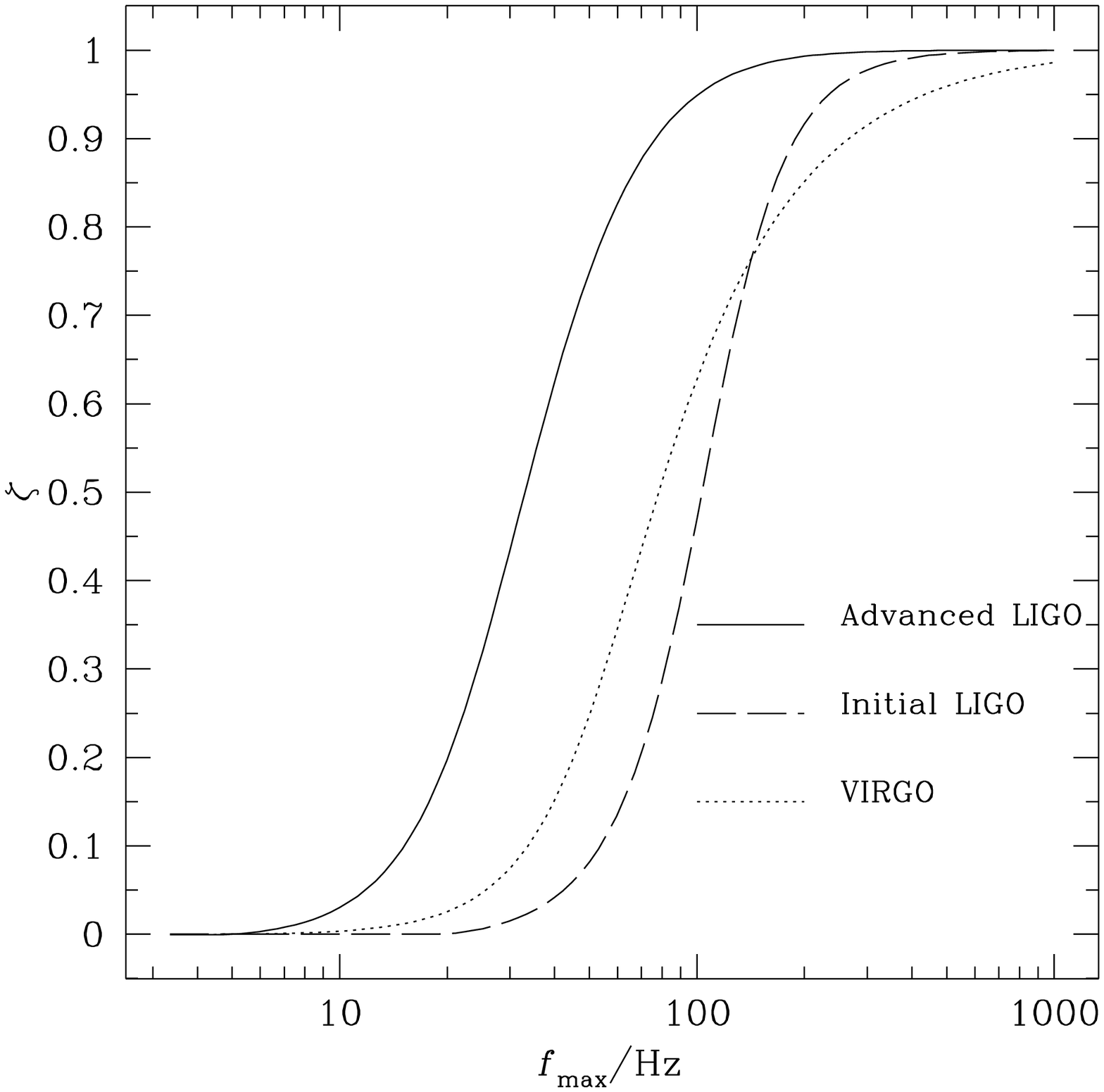}
\caption{The signal-to-noise ratio of the radiation from an
inspiraling compact binary in an interferometric gravitational wave
depends, through the function $\zeta$ (defined in
eq.~\protect{\ref{eqn:zeta}}) on the redshifted orbital frequency
$f_{\text{max}}$ of the system's last orbit before coalescence. Here
is shown $\zeta(f_{\text{max}})$ for initial LIGO (dashed), VIRGO
(dotted) and advanced LIGO (solid) interferometers. For more detail,
see the discussion in section \protect{\ref{sec:ligo/virgo}} and
\protect{\ref{subsec:snr}}.}\label{fig:zeta}
\end{figure}

Figure \ref{fig:zeta} shows $\zeta$ for early LIGO and VIRGO
interferometers and also more advanced LIGO interferometers as have
been discussed in the literature \cite{ligo89a,finn93a}. The solid
curve shows $\zeta$ representative of advanced interferometers, while
the dashed and dotted curves show $\zeta$ representative of initial
LIGO and VIRGO interferometers. As the LIGO detector evolves, $r_0$
will increase from approximately $20\,\text{Mpc}$ toward
$350\,\text{Mpc}$, and $\zeta$ will evolve from the dashed curve
toward the solid curve. Setting aside the overall sensitivity gain
with increasing $r_0$, the evolution of $\zeta$ signifies an
increasing {\em relative} sensitivity to systems with small
$f_{\text{max}}$: {\em i.e.,} systems with larger total masses and/or
redshifts.

VIRGO will consist of a single 3~Km interferometer. The target noise
curve for the initial operation of the VIRGO interferometer is
described in \cite{bradaschia90a}; for this interferometer $r_0$ is
$13\,\text{Mpc}$. Like the LIGO interferometers, incremental
improvements in VIRGO will increase $r_0$ and it is reasonable to
assume that these improvements could raise $r_0$ for VIRGO to
approximately $200\,\text{Mpc}$. The dotted curve in figure
\ref{fig:zeta} shows $\zeta$ for the initial VIRGO
interferometers. Note how, even though $r_0$ is the same for both the
initial LIGO and VIRGO interferometers, the relative sensitivity of
these two detectors to signals at low and high frequency ---
corresponding to more or less massive binaries --- is markedly
different. Improvements in the VIRGO interferometer can be expected to
evolve $\zeta$ toward the solid curve as well. Note that the initial
VIRGO interferometers are expected to be relatively more sensitive to
massive ({\em i.e.,} low $f_{\text{max}}$) binaries than are the
initial LIGO interferometers.

\begin{figure}
\epsfxsize=\columnwidth\epsffile{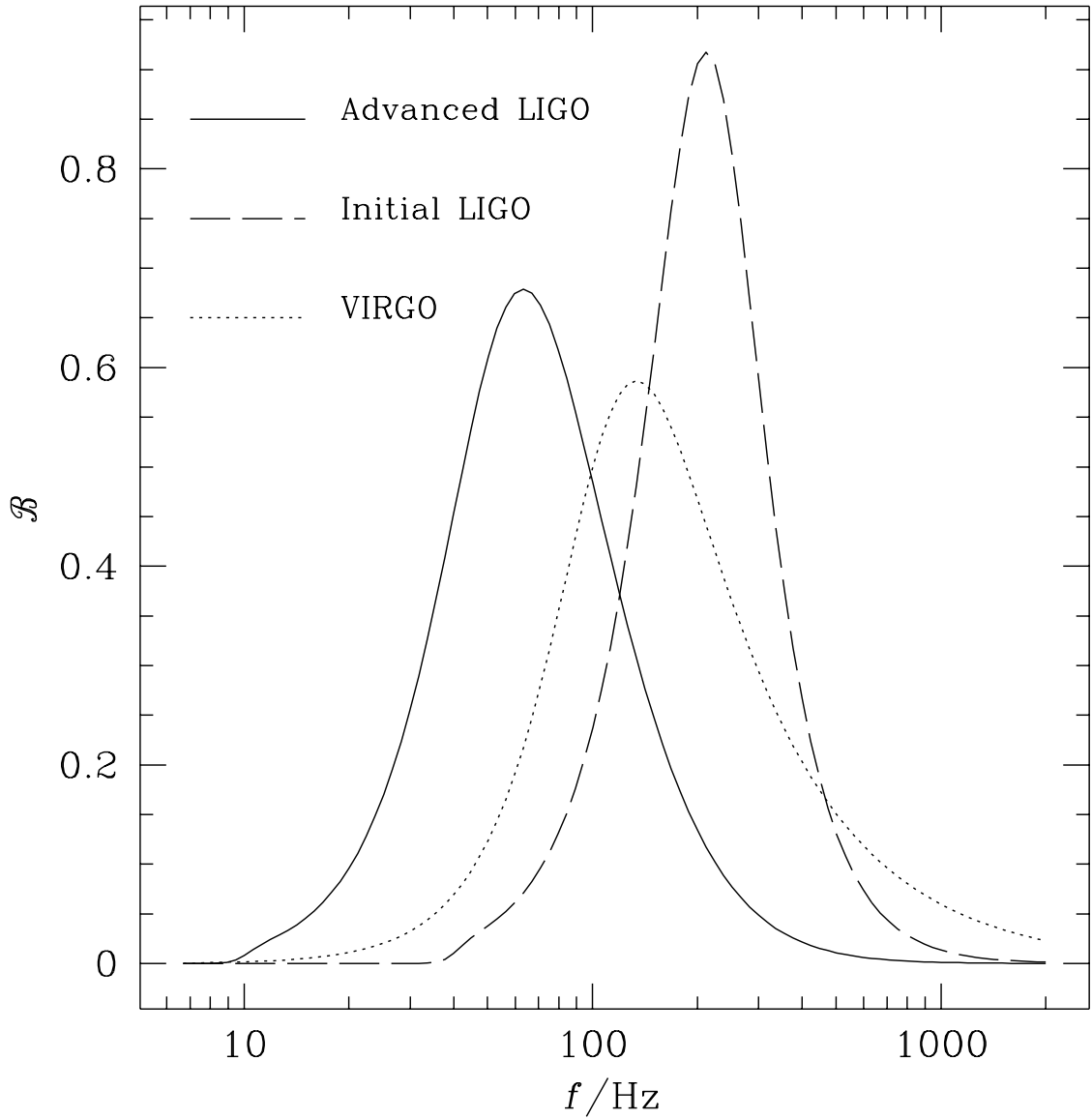}
\caption{The bandwidth function ${\cal B}(f)$ (defined in
eq.~\protect{\ref{eqn:bwidth}}) describes the fraction of $\rho^2$
contributed by signal power at frequency $f$ in a logarithmic
frequency interval. Here is shown $\cal B$ for initial LIGO, VIRGO and
advanced LIGO interferometers. The bandwidth function is especially
useful for determining over what frequency interval an approximate
model of the detector response to the radiation, which might be used
for identifying the presence of a signal in detector output, must
accurately mimic the real detector response. For more details see the
discussion in section \protect{\ref{sec:bandwidth}}.}
\label{fig:bwidth}
\end{figure}

For neutron star binaries $f_{\text{max}}\simeq710\,\text{Hz}$
(eq.~\ref{eqn:fmax}) and $\zeta(f_{\text{max}})\simeq1$ in any of the
proposed LIGO and VIRGO interferometers (see
fig.~\ref{fig:zeta}). Figure \ref{fig:bwidth} shows the bandwidth
function for neutron star binary observations in early VIRGO (dotted
curve) and LIGO (dashed curve) interferometers and also more advanced
LIGO interferometers (solid curve). It is clear that the approximate
detector response models for binary inspiral observations must evolve
with the interferometers: in the early interferometers the detector
response models will need to be quite accurate at high frequencies
($f\simeq200\,\text{Hz}$) but not at low ones; on the other hand, in
more advanced interferometers these models will need to be accurate at
low frequencies ($f\simeq60\,\text{Hz}$) and, to obtain this low
frequency accuracy, the high frequency performance may be sacrificed
at no cost to the signal detectability.

\section{Discussion}\label{sec:discussion}

\subsection{Introduction}

Consider a catalog of binary inspiral events with signal-to-noise
ratio $\rho$ greater than a threshold $\rho_0$. Suppose that for each
event in the catalog $\rho$ and $\cal M$ are known. How do we make use
of the catalog to test cosmological models against cataloged
observations?

The simplest test involves the distribution of events with $\rho$.  As
the catalog limit $\rho_0$ decreases, sources at increasingly larger
distance become members. The number of added sources depends on the
increase of the spatial volume, the density of sources at that
distance, and (since these sources are {\em events} that occur at a
given {\em rate}) the cosmological redshift. Thus, adopting a
cosmological model ${\cal C}$ implies an expected distribution
$P(\rho|\rho_0, {\cal C},{\cal D})$ for the catalog events taken with
detector ${\cal D}$. Denote the cataloged binary inspiral
signal-to-noise ratio observations by $\{\rho|\rho>\rho_0\}$. Suppose
that, before studying these observations, we have reason or are
otherwise prejudiced to believe the probability that $\cal C$ is the
correct cosmological model is $P({\cal C})$. Using Bayes' law of
conditional probabilities, the {\em a posteriori} probability that we
assign to model $\cal C$ after considering the observations is
\begin{equation}
P({\cal C}|\{\rho|\rho>\rho_0\}) \propto P({\cal C})\prod_i
P(\rho_i|\rho_0,{\cal C},{\cal D}_i),
\end{equation}
where $\rho_i$ is the signal-to-noise ratio of the $i^{\text{th}}$
catalog observation and ${\cal D}_i$ represents the detector
configuration ($r_0$, $\zeta$) when the observation was
made\footnote{Changes in the detector noise spectrum change the
relation between $\rho$ and the cosmological model. This is not a
problem in the interpretation of a catalog as long as the detector
properties are properly associated with each cataloged
observation.}. This test is exactly analogous to number-flux
cosmological tests using distant galaxies.

More subtle tests involve other distributions of cataloged events. For
example, each source in the catalog is characterized by its observed
chirp mass $\cal M$. The observed chirp mass depends on both the
intrinsic chirp mass ${\cal M}_0$ and the redshift $z$ (see
eq.~\ref{eqn:M}); consequently, the distribution of events in the
catalog with $\cal M$ depends on the cosmological model. Denote the
cataloged chirp mass observations by $\{{\cal
M}|\rho>\rho_0\}$. Adopting a cosmological model $\cal C$ implies an
expected distribution $P({\cal M}|\rho_0,{\cal C}, {\cal D})$. As
before, if we initially favor model ${\cal C}$ with probability
$P({\cal C})$, then after studying the observations the probability we
ascribe to model ${\cal C}$ is
\begin{equation}
P({\cal C}|\{{\cal M}|\rho>\rho_0\})
\propto P({\cal C})\prod_i 
P({\cal M}_i|\rho_0,{\cal C},{\cal D}_i),
\end{equation}
where ${\cal M}_i$ represents the $i^{\text{th}}$ cataloged chirp mass
observation. 

Of the four parameters that describe, at quadrupole order, binary
inspiral observed in a single interferometer, only $\rho$ and $\cal M$
convey astrophysically interesting information. The distributions
$P(\rho|\rho_0,{\cal C},{\cal D})$ and $P({\cal M}|\rho_0,{\cal
C},{\cal D})$ are each
integrals over the distribution $P(\rho,{\cal M}|\rho_0,{\cal C},{\cal D})$
that completely characterizes the catalog:
\begin{mathletters}
\begin{eqnarray}
P(\rho|\rho_0,{\cal C},{\cal D}) &=& \int P(\rho,{\cal
M}|\rho_0,{\cal C},{\cal D}) d{\cal M}\\
P({\cal M}|\rho_0,{\cal C},{\cal D}) &=& \int_{\rho_0}^\infty P(\rho,{\cal
M}|\rho_0,{\cal C},{\cal D}) d\rho.
\end{eqnarray}
\end{mathletters}
These integrals are {\em summaries} of the catalog contents: as such,
they are less informative than $P(\rho,{\cal M}|\rho_0,{\cal C},{\cal
D})$. The most sensitive test that we can make using the catalog
involves not a summary, but the full information available: given the
observations $\{\rho,{\cal M}|\rho>\rho_0\}$, the probability that
cosmological model ${\cal C}$ describes our universe is
\begin{equation}
P({\cal C}|\{\rho,{\cal M}\},\rho_0) \propto
P({\cal C})\prod_i P(\rho_i,{\cal M}_i|\rho_0,{\cal C},{\cal D}_i).
\end{equation}

Cosmological tests based on the summary distributions are still useful
to make. Summary distribution often depend only weakly or not at all
on some of the parameters of the model ${\cal C}$; in that case, the
effective dimensionality of ${\cal C}$ is reduced in the summary test
and we may be able to distinguish more closely among cosmological
models described by the remaining parameters than with a test using
the full distribution. This is particularly true when the number of
observations in the catalog is small.

\subsection{Outline}

In this section I give general expressions for the distributions
$P(\rho,{\cal M}|\rho_0,{\cal C},{\cal D})$ and the summary
distributions $P(\rho|\rho_0,{\cal C},{\cal D})$ and $P({\cal
M}|\rho_0,{\cal C},{\cal D})$ that can be evaluated in any homogeneous
and isotropic cosmological model; in addition, I evaluate the summary
distributions explicitly for matter-dominated
Friedmann-Robertson-Walker (FRW) models. The essential properties of
FRW cosmological models are summarized in section~\ref{sec:frw}.

In addition to the distributions $P(\rho,{\cal M}|\rho_0,{\cal
C},{\cal D})$,
$P(\rho|\rho_0,{\cal C},{\cal D})$ and $P({\cal M}|\rho_0,{\cal
C},{\cal D})$ are other
catalog properties of intrinsic interest. In a given cosmological
model there is a distance beyond which no binary inspiral of fixed
intrinsic chirp mass will have signal-to-noise ratio greater than the
catalog limit $\rho_0$, and I calculate this {\em catalog depth} $z_0$
in section \ref{sec:depth}. A catalog takes time to build, and, in a
given period of time, the size of a binary inspiral catalog is limited
by the rate $dN/dt$ at which binaries coalesce with $\rho$ greater
than the catalog threshold $\rho_0$. In section \ref{sec:rate} I
calculate both the expected rate of binary inspiral observations and
the distribution $P(\rho|\rho_0,{\cal C},{\cal D})$ for advanced LIGO
interferometers and describe how these depend on the cosmological
model ${\cal C}$. Finally, in section \ref{sec:spectrum} I calculate
the distribution $P({\cal M}|\rho_0,{\cal C},{\cal D})$ and describe how it
depends on ${\cal C}$.

\subsection{Cosmological model}\label{sec:frw}

Specific examples of the catalog distributions, catalog depth, and
rate of binary inspiral observations made in this section
are in the context of matter-dominated Friedmann-Robertson-Walker (FRW)
cosmological models. The Robertson-Walker spacetime metric has the
line element
\begin{equation}
ds^2 = - dt^2 + a^2(t)\left[{dr^2\over 1-kr^2} + r^2 \left(d\theta^2 +
\sin^2\theta d\phi^2\right) \right],
\end{equation}
where $a(t)$ is the usual (dimensioned) scale factor, $r$ is a
dimensionless parameter related to the area of spheres of constant
radius, and $k$ is $+1$, $-1$, or $0$ depending on whether the spatial
geometry of the slices of homogeneity has positive, negative, or zero
curvature ({\em i.e.,} is closed, open, or flat). In matter dominated
FRW models (which have vanishing cosmological constant) the co-moving
radial coordinate $r$ and the luminosity distance $d_L$ can be written
in terms of the redshift explicitly:
\begin{mathletters}
\label{eqn:frw}
\begin{eqnarray}
r &=& {z q_0 +
\left(q_0-1\right)\left[\left(1+2q_0z\right)^{1/2}-1\right]\over 
a_0 H_0 q_0^2\left(1+z\right)}, \label{eqn:r} \\
d_L &=& a_0\left(1+z\right) r, \label{eqn:dL}
\end{eqnarray}
\end{mathletters}
where $a_0$, $H_0$, and $q_0$ are the scale factor, expansion rate,
and curvature parameter (deceleration parameter) at the present epoch
\cite[eq.~15.3.24]{weinberg72a}. 

\subsection{Sample depth}\label{sec:depth}

The signal-to-noise ratio of an inspiraling neutron star binary system
with intrinsic chirp mass ${\cal M}_0$ is
\begin{equation}
\rho = 8 \Theta\left(r_0\over d_L\right)
\left({\cal M}_0\over 1.2\text{M}_\odot\right)^{5/6}
\left(1+z\right)^{5/6} \zeta(f_{\text{max}}).  
\end{equation}
Since $\Theta$ is between $0$ and $4$, even an optimally oriented
binary system has $\rho$ less than $\rho_0$ when $z$ is greater than
$z_0$, where $z_0$ satisfies
\begin{equation}
4 = {\rho_0d_L(z_0)\over8r_0\zeta(f_{\text{max}})^{5/6}} 
\left(1.2\,\text{M}_\odot\over{\cal M}_0\right)^{5/6}.
\label{eqn:z0implicit}
\end{equation}
Evaluation of $z_0$ requires $\zeta(f_{\text{max}})$, which depends on
the details of the detector's noise power spectral density (through
$\zeta$) as well as the binary's component masses and redshift
(through $f_{\text{max}}$). For advanced interferometers
$\zeta\simeq1$ as long as $f_{\text{max}}\gtrsim70\,\text{Hz}$ (see
fig. \ref{fig:zeta}) while
$f_{\text{max}}\simeq710\,\text{Hz}\left[2.8\,\text{M}_\odot/(1+z_0)
M\right]$ for symmetric binary inspiral (see eq. \ref{eqn:fmax}).
Consequently, we can approximate $\zeta\simeq1$ as long as
$1+z_0\lesssim10\left(2.8\,\text{M}_\odot/M\right)$. For small $z_0$
we can approximate $d_L\simeq z/H_0$; then
\begin{mathletters}
\begin{eqnarray}
z_0 &\simeq& {32 H_0 r_0\over\rho_0}\left({\cal
M}_0\over1.2\,\text{M}_\odot\right)^{5/6}\\
&\simeq& 0.48h\left(8\over\rho_0\right)
\left(r_0\over355\,\text{Mpc}\right) 
\left({\cal M}_0\over 1.2\,\text{M}_\odot\right)^{5/6},
\end{eqnarray}
\end{mathletters}
which is much less than $10$. Thus, the approximation $\zeta\simeq1$
is a good one for binary systems with solar mass components, but not
for binaries whose components are on order $5\,\text{M}_\odot$.

For a specific example, focus attention first on solar mass component
binaries. Then $\zeta\simeq1$ for any of the proposed LIGO or VIRGO
interferometers (at the end of this subsection I return to consider
briefly the case of more massive binary systems where this is not
true). In an Einstein-deSitter ($q_0=1/2$) cosmological model $z_0$ is
then given explicitly by
\begin{mathletters}
\label{eqn:z0EdS}
\begin{equation}
z_0 = \left[\beta + {256\over9\beta}\alpha_0^2
+ {16\over3}\alpha_0\right]^6-1,
\end{equation}
where
\begin{eqnarray}
\alpha_0 &=& {8H_0r_0\over\rho_0}\left({\cal
M}_0\over1.2\,\text{M}_\odot\right)^{5/6}\label{eqn:alpha0}\\
&=& 0.12h\left(8\over\rho_0\right)
\left(r_0\over355\,\text{Mpc}\right) 
\left({\cal M}_0\over 1.2\,\text{M}_\odot\right)^{5/6},
\end{eqnarray}
$h$ is the Hubble parameter ($H_0$ in units of $100\,\text{Km/s/Mpc}$),
and
\begin{equation}
\beta = \left[{1\over2}+\left(16\alpha_0\over3\right)^3
+\left({1\over4}+{4096\over27}\alpha_0^3\right)^{1/2}\right]^{1/3}.
\end{equation}
\end{mathletters}
More generally ($q_0\neq1/2$ but $\zeta$ still unity) $z_0$ satisfies
\begin{mathletters}
\label{eqn:z0}
\begin{equation}
z_0 = x^6-1
\end{equation}
where $x$ is a root of 
\begin{eqnarray}
0 &=& x^{12} + 8 q_0 \alpha_0 x^{11} + 16 q_0^2\alpha_0^2 x^{10}\nonumber\\
&&\qquad{} - 2 q_0 x^6 - 8\alpha_0(2q_0-1) x^5 + 2q_0-1.
\end{eqnarray}
\end{mathletters}
The appropriate root of this equation can be found as a power series
for small $\alpha_0$: 
\widetext
\begin{eqnarray}
z_0 &=& 4\alpha_0 
+ {8\over3}\left(3q_0+2\right)\alpha_0^2
+ 8\left(6q_0-1\right)\alpha_0^3\nonumber\\
&&\qquad{}
+ {160\over81} \left(54q_0^2 + 9 q_0 - 7\right)\alpha_0^4
+ {5600\over243}\left(27q_0^2-{447\over25}q_0+{323\over100}\right)\alpha_0^5
+ {\cal O}(\alpha_0^6)
\label{eqn:z0ps}
\end{eqnarray}
This truncated expansion is accurate to better than 0.2\% for
$\alpha_0<0.12$ and $0\leq q_0\leq1$; for $\alpha_0<0.1$ and
$0\lesssim q_0\lesssim 3/4$ the accuracy is better than 0.01\%.  An
asymptotic expansion for $z_0$ valid for large $\alpha_0$ (but $\zeta$
still $1$) is
\begin{eqnarray}
z_0 &=& 4096 q_0^6 \alpha_0^6 - 384
\sqrt{2q_0^5}\left(q_0-1\right)\alpha_0^3
+ 6q_0-1 + {\cal O}\left(\alpha_0^{-3}\right).
\end{eqnarray}
This asymptotic expansion is accurate to better than 1\% for
$\alpha_0\gtrsim1/2$ and $0.1\lesssim q_0\lesssim 1$.

\begin{figure}
\epsfxsize=\columnwidth \epsffile{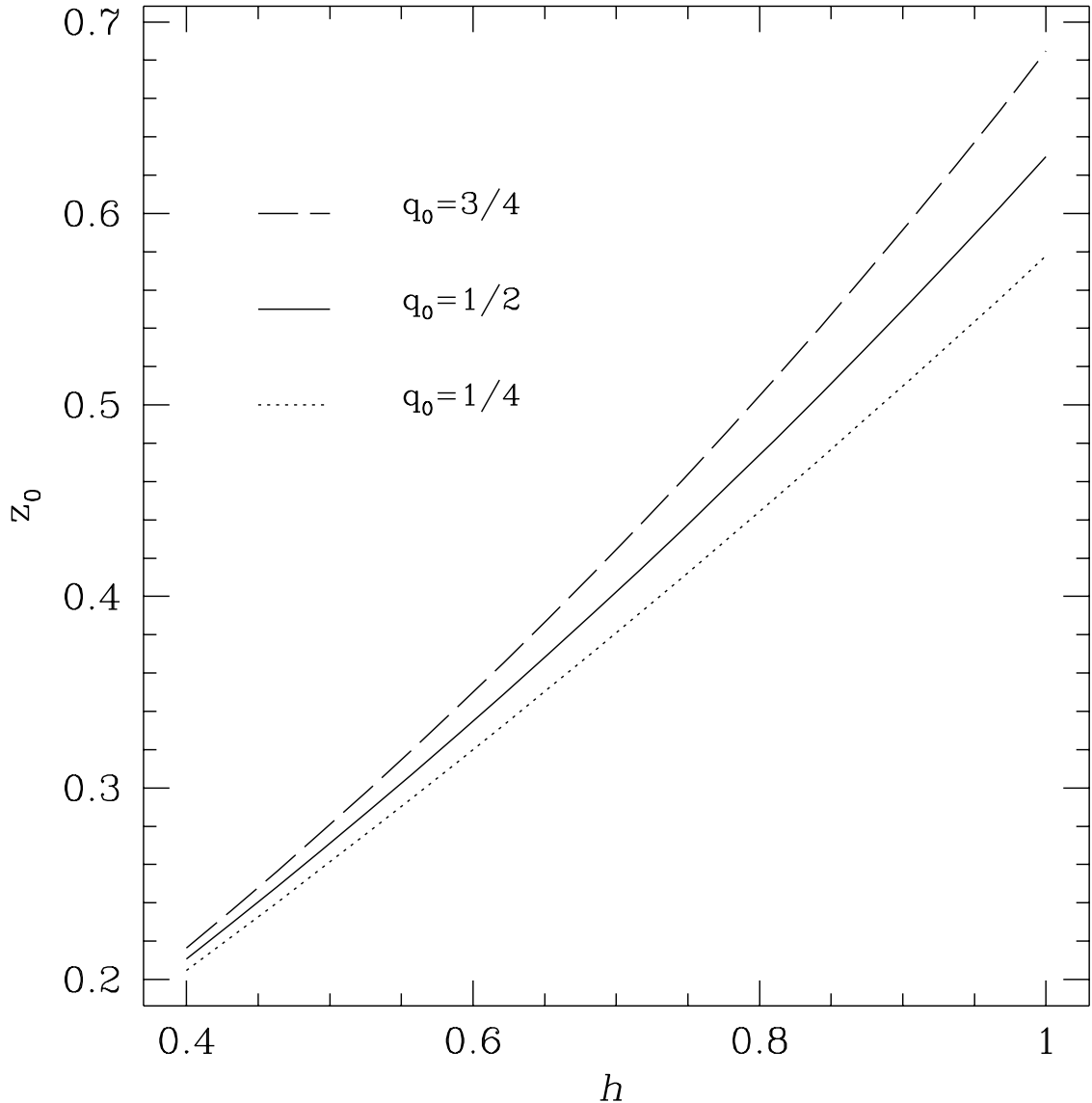}
\caption{The distance to the farthest inspiraling binary system with
signal-to-noise ratio $\rho$ greater than a threshold $\rho_0$ depends
on the detector noise spectrum, the binary system component masses,
and the cosmological model. Shown here is the redshift to the farthest
ns-ns binary system observable with $\rho\geq8$ in an advanced LIGO
detector as a function of the Hubble parameter $h$ (the Hubble
constant in units of $100\,\text{Km/s/Mpc}$).  The three curves
represent matter dominated Friedmann-Robertson-Walker cosmological
models with different $q_0$: a closed model with $q_0=3/4$ (dashed
curve), a flat model ($q_0=1/2$, solid curve), and an open model with
$q_0=1/4$ (dotted curve). For more discussion see
\protect{\ref{sec:depth}}.}\label{fig:z0short}
\end{figure}

\narrowtext 
Figure \ref{fig:z0short} shows $z_0$ as a function of $h$ for neutron
star binaries with ${\cal M}_0=1.19\,\text{M}_\odot$ (corresponding to
$1.37\,\text{M}_\odot$ neutron stars), $\rho_0=8$, $r_0=355$, and
three different values of $q_0$ corresponding to an open ($q_0=1/4$,
dotted curve), flat ($q_0=1/2$, solid curve) and closed ($q_0=3/4$,
dashed curve) cosmological model.  The redshift of the most distant
such source is less than $1/2$ for $h<0.8$ and does not exceed $7/10$
for $h<1$. The sensitivity of $z_0$ to $q_0$ is modest but
significant: inspection of equation \ref{eqn:z0} shows that
\begin{equation}
z_0 \simeq 4\alpha_0\left[1+ \alpha_0\left(2q_0-1\right) \right]
+ {\cal O}\left[\alpha_0^3,\left(2q_0-1\right)^2\right],
\end{equation}
where $\alpha_0\simeq0.12h$ for advanced LIGO interferometers. For
open spatial geometries ($2q_0-1<0$) the most distant sources are at
smaller redshifts than in closed spatial geometries (where
$2q_0-1>0$). 

\begin{figure}
\epsfxsize=\columnwidth\epsffile{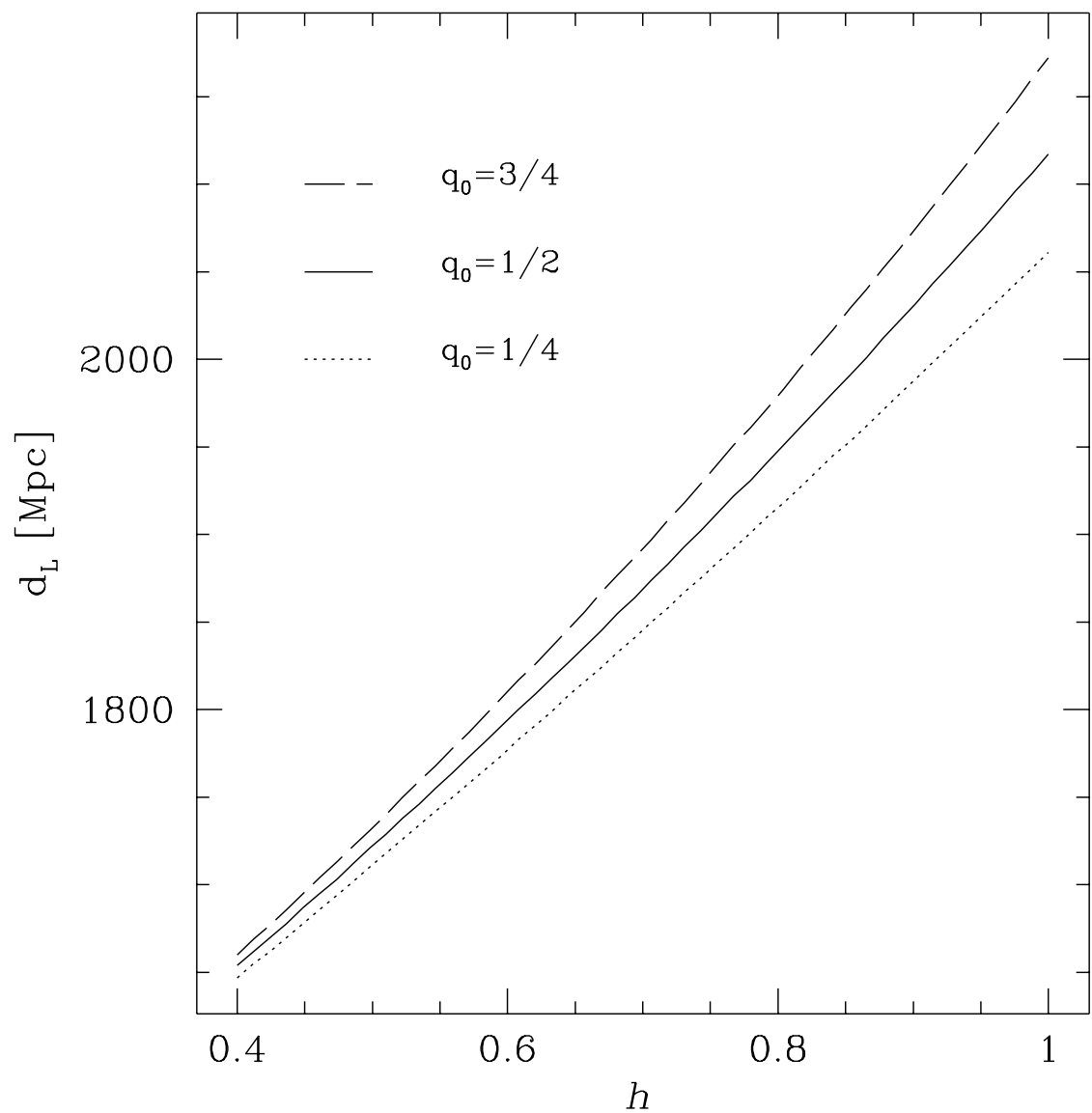}
\caption{The same as figure \protect{\ref{fig:z0short}}, except shown
here is the luminosity distance instead of the
redshift.}\label{fig:dl}
\end{figure}

\widetext
In luminosity distance, the sample depth is
\begin{eqnarray}
d_{L,0} &=& {4\alpha_0\over H_0}\left[
1 
+ {40\over3}\alpha_0 
+ {40\over 3}\left(2q_0+1\right)\alpha_0^2 
+ {1280\over81}\left(9q_0-2\right)\alpha_0^3 
\right.\nonumber\\&&\qquad\left.{}
+ {40\over243}\left(2052q_0^2 - 180q_0 - 143\right)\alpha_0^4
+ {\cal O}\left(\alpha_0^6\right)
\right].
\end{eqnarray}
Figure \ref{fig:dl} shows $d_{L,0}$ for the same cases ($h$, $q_0$) as
figure \ref{fig:z0short} shows $z_0$. Advanced LIGO interferometers
may observe neutron star binaries with $\rho$ greater than $8$ at
luminosity distances of order 2~Gpc.

\begin{figure}
\epsfxsize=\columnwidth\epsffile{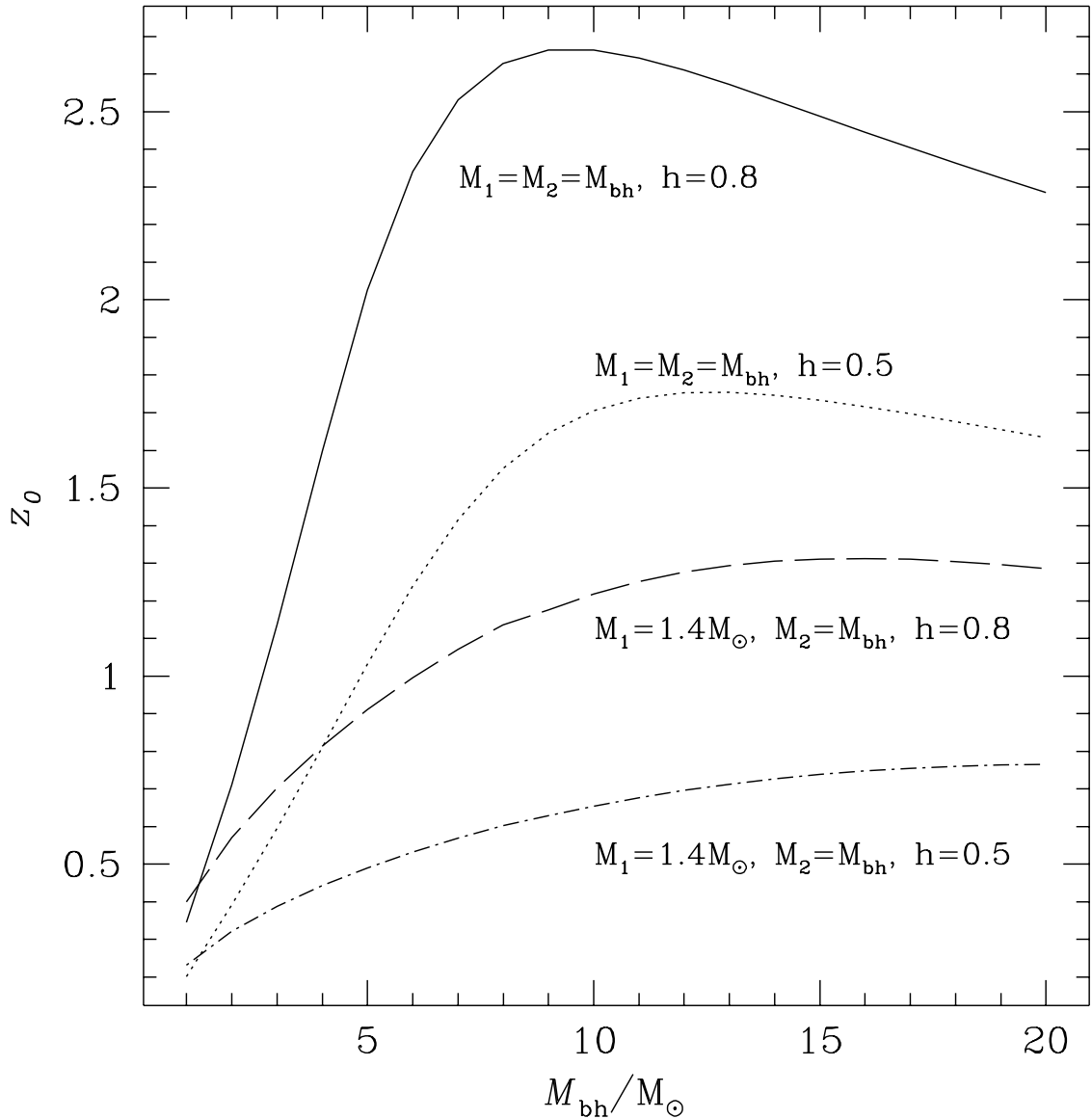}
\caption{The expected redshift to the farthest inspiraling binary
system observed by an advanced LIGO detector with $\rho\geq8$ in
matter-dominated Einstein-deSitter cosmological models
($q_0=1/2$). Results for symmetric binary systems, consisting of two
components each of mass $M_{\text{bh}}$, and asymmetric binary
systems, consisting of a $1.4\,\text{M}_\odot$ component and a
$M_{\text{bh}}$ component, are shown for $h=0.8$ and $h=0.5$. The
maximum observable depth at any signal-to-noise threshold is limited
by the cosmological model and the properties of the detector; for
advanced LIGO interferometers and $\rho_0=8$ it peaks for symmetric
binaries composed of $10\,\text{M}_\odot$ black holes. For more
discussion see section \protect{\ref{sec:depth}}.}\label{fig:z0long}
\end{figure}

\narrowtext For more massive binaries $\zeta(f_{\text{max}})$ is
substantially less than unity at $z_0$ and the approximation
$\zeta\simeq1$ is no longer valid. Figure \ref{fig:z0long} shows $z_0$
(given implicitly by eq.~\ref{eqn:z0implicit}) for an advanced LIGO
detector ($r_0=355\,\text{Mpc}$) and a range of binary systems in two
distinct Einstein-de Sitter ($q_0=1/2$) cosmological models. One pair
of curves show $z_0$ for symmetric binaries consisting of two ({\em
e.g.}) black holes each of mass $M_{\text{bh}}$, while the second pair
of curves shows $z_0$ for asymmetric binaries, consisting of ({\em
e.g.,}) a black hole of mass $M_{\text{bh}}$ and a neutron star of
mass $1.4\,\text{M}_\odot$. Note how, in {\em all} cases, $z_0$ does
not increase without bound as $M_{\text{bh}}$ increases; instead, it
has a maximum value for some $M_{\text{bh}}$ and decreases as
$M_{\text{bh}}$ increases further.

The finite bandwidth of the detector and its overlap with the signal
power, as represented by $\zeta$, determines the maximum redshift at
which a detector can observe binary inspiral. If $\zeta$ were constant
then $z_0$ would increase monotonically with ${\cal M}_0$ (see
eq.~\ref{eqn:z0implicit}); however, $\zeta$ is not constant and as
$\cal M$ increases $f_{\text{max}}$ decreases both because
$f_{\text{ICO}}$ is decreasing and also because $z_0$ is increasing
($f_{\text{max}}\propto M/(1+z)$; see eq.~\ref{eqn:fmax}). Eventually
$\zeta$ begins to decrease with increasing ${\cal M}_0$ or decreasing
$f_{\text{max}}$ (for advanced LIGO interferometers, $\zeta$ remains
unity until $f_{\text{max}}\simeq70\,\text{Hz}$; see
fig.~\ref{fig:zeta}). Once $\zeta$ begins to decrease so to will
$z_0$. For sufficiently large ${\cal M}_0$ even a local binary ($z=0$)
will coalesce before the gravitational radiation signal enters the
detector bandwidth.

Thus, decreasing the detector noise at low frequencies has a dual
effect: by increasing $r_0$ it increases the overall detector
sensitivity, and by decreasing the frequency below which $\zeta\ll1$
it further boosts the sensitivity of the interferometer to more
massive and more distant binary systems. Advanced LIGO detectors are
expected to be most sensitive to inspiraling binaries with two
$10\,\text{M}_\odot$ black hole components, which will be observable
with $\rho$ greater than 8 at redshifts greater than 1.5.

\subsection{Coalescence rate above threshold}\label{sec:rate}

The distribution of catalog events with $\rho$ is given by 
\begin{mathletters}
\label{eqn:P(rho|I)}
\begin{equation}
P(\rho|\rho_0,{\cal C},{\cal D}) = \left\{\begin{array}{ll}
{{d^2N/dt\,d\rho}\over dN/dt}&\rho>\rho_0,\\
0&\rho<\rho_0,
\end{array}\right.
\end{equation}
where
\begin{eqnarray}
{d^2N\over dt\,d\rho} &=& \int d{\cal M}_0 {d^3N\over dt\,d\rho\,d{\cal M}_0},\\
{d^3N\over dt\,d\rho\,d{\cal M}_0} &=& 
\int dz{dr\over dz}{4\pi a_0^3 r^2\over\sqrt{1-kr^2}}
{\dot{n}_0{\cal E}(z)\over1+z}\nonumber\\
&&\qquad\!\times{{\cal P}\left({{\cal M}_0}|z\right)\over1+z}
P_\rho(\rho|z,{\cal M}_0,{\cal C},{\cal D}),\\
{dN\over dt} &=& \int_{\rho_0}^\infty d\rho\,{d^2N\over
dt\,d\rho},
\label{eqn:dN/dt'} 
\end{eqnarray}
\end{mathletters}
and $P_\rho$ is given in equation \ref{eqn:p(rho)}. 

Since $d^2N/dt\,d\rho$ depends on $\rho$ only through $P_\rho$, the
integral over $\rho$ in equation \ref{eqn:dN/dt'} can be evaluated
explicitly. First note that
\begin{mathletters}
\begin{eqnarray}
\int_{\rho_0}^\infty d\rho\,P_\rho(\rho|{\cal M}_0,z,{\cal C},{\cal D}) &=&
\int_x^\infty P_\Theta(\Theta) \\ 
&\equiv& C_\Theta(x),
\end{eqnarray}
where
\begin{equation}
x = {\rho_0\over8}{d_L\over r_0\left(1+z\right)^{5/6}}
\left(1.2\,\text{M}_\odot\over{\cal M}_0\right)^{5/6}. 
\label{eqn:x-cum}
\end{equation}
\end{mathletters}
The probability density $P_\rho$ is a conditional one: it depends on
${\cal M}_0$, $z$, and the cosmological model. In contrast, the
distribution $P_\Theta$ of $\Theta$ is universal: it is independent of
all the specific properties of the binary or the detector.
$C_\Theta(x)$ is universal in this same way and can be evaluated with
$P_\Theta$.  In terms of $C_\Theta$ the total rate of inspirals with
$\rho$ greater than $\rho_0$ is
\begin{mathletters}
\label{eqn:dN/dt}
\begin{equation}
{dN\over dt} = \int d{\cal M}_0 {d^2N\over dt\, d{\cal M}_0},
\end{equation}
where
\begin{eqnarray}
{d^2N\over dt\,d{\cal M}_0} &=& 
\int_0^\infty dz\,
{dr\over dz}
{4\pi a_0^3 r^2\over\sqrt{1-kr^2}}
{\dot{n}_0{\cal E}(z)\over1+z}
{\cal P}\left({\cal M}_0|z\right) 
\nonumber\\
&&\qquad{}\times 
C_\Theta
\left[d_L(z)\over\alpha_0\left(1+z\right)^{5/6}\zeta(f_{\text{max}})\right]
\label{eqn:dN/dtdM0}
\end{eqnarray}
and 
\begin{equation}
\alpha_0 \equiv {8 H_0 r_0\over\rho_0}\left({\cal M}_0\over
1.2\,\text{M}_\odot\right)^{5/6}.
\end{equation}
\end{mathletters}
Equation \ref{eqn:dN/dt} for $dN/dt$, which is also the total rate of
binary inspiral with $\rho>\rho_0$, and equations \ref{eqn:P(rho|I)} for
$P(\rho|\rho_0,{\cal C},{\cal D})$ are valid for any homogeneous and
isotropic cosmological model.

\subsubsection{Examples: Constant ${\cal M}_0$}\label{sec:constM0}

Turn now to the specific example of neutron star binaries in
matter-dominated FRW cosmological models, so that $\zeta\simeq1$ and
$r$ and $d_L$ are given by equations \ref{eqn:frw}. Assume also that
evolution in the binary population is unimportant over redshifts
$z\lesssim1/2$; then ${\cal E}$ is unity and ${\cal P}$ is independent
of $z$. Finally, use the approximation given above for $P_\Theta$
(eq.~\ref{eqn:pTheta}) to evaluate $C_\Theta$:
\begin{eqnarray}
C_\Theta(x) &\simeq& \left\{
\begin{array}{ll}
1&\text{if $x\leq0$}\\
(1+x)(4-x)^4/256&\text{if $0\leq x\leq4$}\\
0&\text{if $4<x$.}
\end{array}
\right.\label{eqn:c(x)}
\end{eqnarray}
\widetext
The integral for $d^2N/dt\,d{\cal M}_0$ (eq.~\ref{eqn:dN/dtdM0}) can
then be evaluated as a power series in small $\alpha_0$:
\begin{mathletters}
\label{eqn:approx-dN/dtdM0}
\begin{equation}
{d^2N\over dt\,d{\cal M}_0} = 
\left(dN\over dt\right)_0{\cal P}({\cal M}_0)
\end{equation}
where
\begin{eqnarray}
\left({dN\over dt}\right)_0 &=&
\left({128\over21}\pi r_0^3\dot{n}_0\right)
\left(8\over\rho_0\right)^3
\left({\cal M}_0\over1.2\text{M}_\odot\right)^{5/2} 
\left[1 
- {25\over9}\alpha_0 
+ {2\over45}\left(6q_0+85\right)\alpha_0^2\right.\nonumber\\
&&\qquad\left.{}
- {224\over33}q_0\alpha_0^3
+ {20\over8019}\left(1188q_0^2 + 9108q_0 - 4123\right)\alpha_0^4
+ {\cal O}\left(\alpha_0^5\right)
\right]\label{eqn:(dN/dt)0}
\end{eqnarray}
\end{mathletters}
and $\alpha_0$ is given by equation~\ref{eqn:alpha0}. 
Similarly, $d^3N/dt\,d\rho\,d{\cal
M}_0$ is given by 
\begin{mathletters}
\label{eqn:approx-dN/dtdrhodM0}
\begin{equation}
{d^3N\over dt\,d\rho\,d{\cal M}_0} = 
\left(d^2N\over dt\,d\rho\right)_0{\cal P}({\cal M}_0)
\end{equation}
where
\begin{eqnarray}
\left({d^2N\over dt\,d\rho}\right)_0 &=&
\left({16\over7}\pi r_0^3\dot{n}_0\right)
\left(8\over\rho\right)^4
\left({\cal M}_0\over1.2\text{M}_\odot\right)^{5/2} 
\left[1 
- {100\over27}\alpha
+ {2\over27}\left(6q_0+85\right)\alpha^2\right.\nonumber\\
&&\qquad\left.{}
- {448\over33}q_0\alpha^3
+ {140\over24057}\left(1188q_0^2 + 9108q_0 - 4123\right)\alpha^4
+ {\cal O}\left(\alpha^5\right)
\right]
\end{eqnarray}
and
\begin{eqnarray}
\alpha &=&
{8H_0 r_0\over\rho}\left({\cal M}_0\over
1.2\,\text{M}_\odot\right)^{5/6}
\nonumber\\
&=&
\alpha_0\left(\rho_0\over\rho\right). 
\label{eqn:alpha}
\end{eqnarray}
\end{mathletters}
Equations \ref{eqn:approx-dN/dtdM0} and \ref{eqn:approx-dN/dtdrhodM0}
for $d^2N/dt\,d{\cal M}_0$ and $d^3N/dt\,d\rho\,d{\cal M}_0$ can be
integrated over ${\cal M}_0$ to find $dN/dt$ and $d^2N/dt\,d\rho$, and
thus $P(\rho|\rho_0,{\cal C},{\cal D})$.

Observational and theoretical evidence suggests that the ns-ns binary
intrinsic chirp mass distribution is sharply peaked (see
sec.~\ref{sec:NConstraints}). In the most extreme limit that all
binary systems have intrinsic chirp mass ${\cal M}_0$, $\cal P$ is a
$\delta$-function in ${\cal M}_0$ and
\begin{mathletters}
\begin{eqnarray}
{dN\over dt} &=& \left(dN\over dt\right)_0\\
P(\rho|\rho_0,{\cal C},{\cal D}) &=& {(d^2N/dt\,d\rho)_0\over(dN/dt)_0} 
\nonumber\\
&=& {3\over\rho}\left(\rho_0\over\rho\right)^3
{
\begin{array}{l}
1 
- {100\over27}\alpha
+ {2\over27}\left(6q_0+85\right)\alpha^2
- {448\over33}q_0\alpha^3\\
\qquad{}
+ {140\over24057}\left(1188q_0^2 + 9108q_0 - 4123\right)\alpha^4
\end{array}
\over
\begin{array}{l}
1 
- {25\over9}\alpha_0 
+ {2\over45}\left(6q_0+85\right)\alpha_0^2
- {224\over33}q_0\alpha_0^3\\
\qquad{}
+ {20\over8019}\left(1188q_0^2 + 9108q_0 - 4123\right)\alpha_0^4
\end{array}
}
+ {\cal O}\left(\alpha_0^5, \alpha^5\right).
\label{P(rho|FRW)}
\end{eqnarray}
\end{mathletters}

\begin{figure}
\epsfxsize=\columnwidth\epsffile{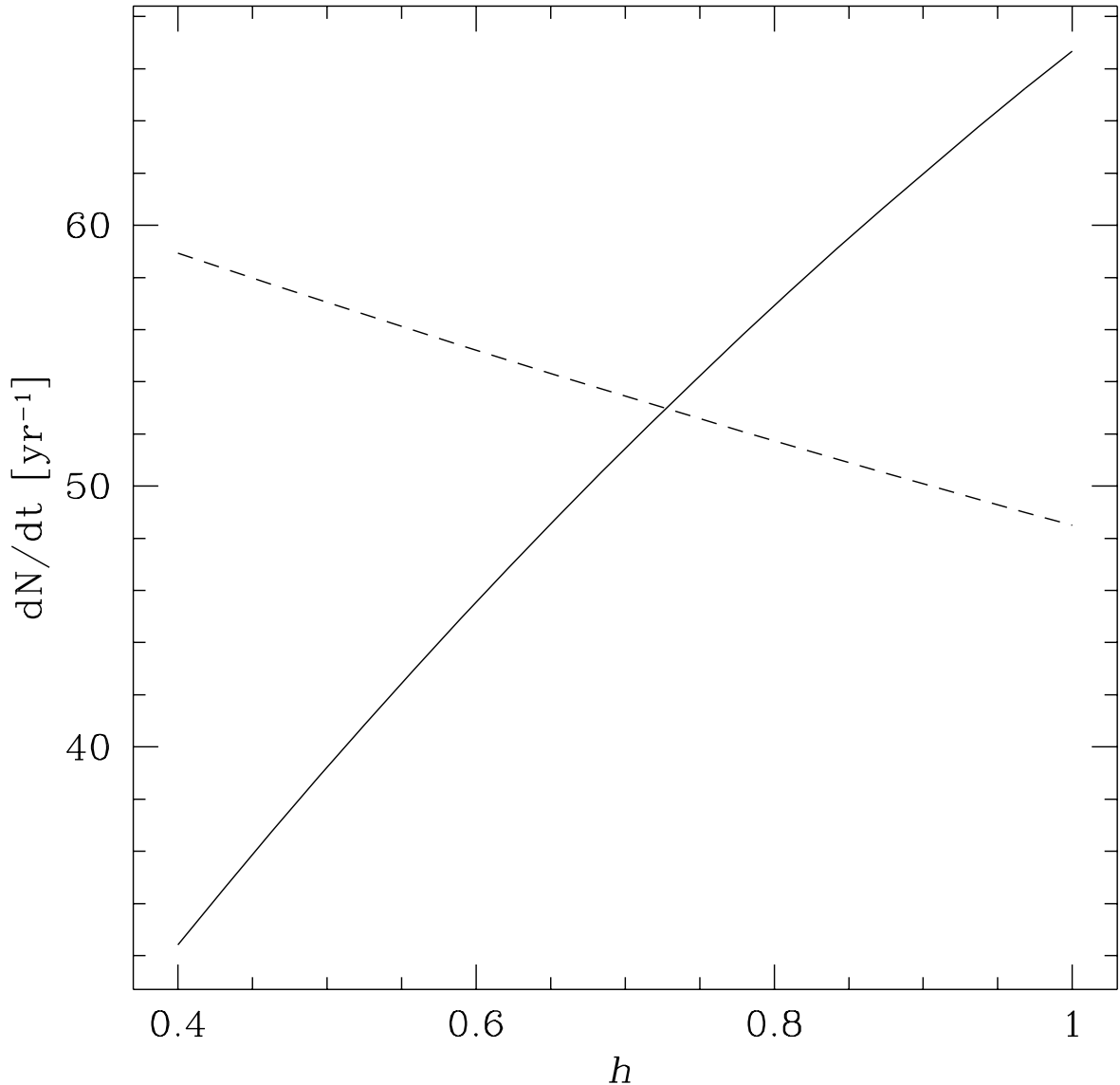}
\caption{The rate of ns-ns binary inspiral observations with
signal-to-noise ratio greater than $8$ in an advanced LIGO detector is
largely insensitive to the neutron star mass range or the deceleration
parameter in matter dominated Friedmann-Robertson-Walker cosmological
models. The solid curve shows the expected rate in an
Einstein-deSitter model as a function of the Hubble parameter $h$
assuming the co-moving ns-ns binary coalescence rate density at the
current epoch is $1.1h\,\text{Mpc}^{-3}\,\text{yr}^{-1}$ (solid
curve); the dashed curve shows the same assuming the rate density is
$8\times10^{-8}\,\text{Mpc}^{-3}\,\text{yr}^{-1}$, which is
independent of $h$. For more discussion see section
\protect{\ref{sec:rate}}.}
\label{fig:dN/dt}
\end{figure}

\narrowtext
Figure \ref{fig:dN/dt} shows $(dN/dt)_0$ as a function of $h$ in an
Einstein-deSitter cosmological model ($q_0=1/2$) for a reasonable
signal-to-noise ratio limit in an advanced LIGO detector ($\rho_0=8$
and $r_0=355\,\text{Mpc}$), typical neutron star masses (${\cal
M}_0=1.19\,\text{M}_\odot$), and two different coalescence rate
densities. The solid curve shows $(dN/dt)_0$ for the ``best guess''
coalescence rate density,
$\dot{n}_0=1.1\times10^{-7}h\,\text{Mpc}^{-3}\,\text{s}^{-1}$ (see
sec.~\ref{sec:n0}), which is proportional to $h$; the dashed curve
shows $(dN/dt)_0$ for a constant coalescence rate density
$\dot{n}_0=8\times10^{-8}\,\text{Mpc}^{-1}\,\text{yr}^{-1}$. The
dashed curve shows how $(dN/dt)_0$ scales for constant $\dot{n}_0$ in
advanced interferometers with interesting $h$. Present estimates of
$\dot{n}_0$ suggest that, if $h=0.75$, advanced LIGO detectors can
expect to observe a little more than 1 neutron star binary inspiral
event per week.

\begin{figure}
\epsfxsize=\columnwidth \epsffile{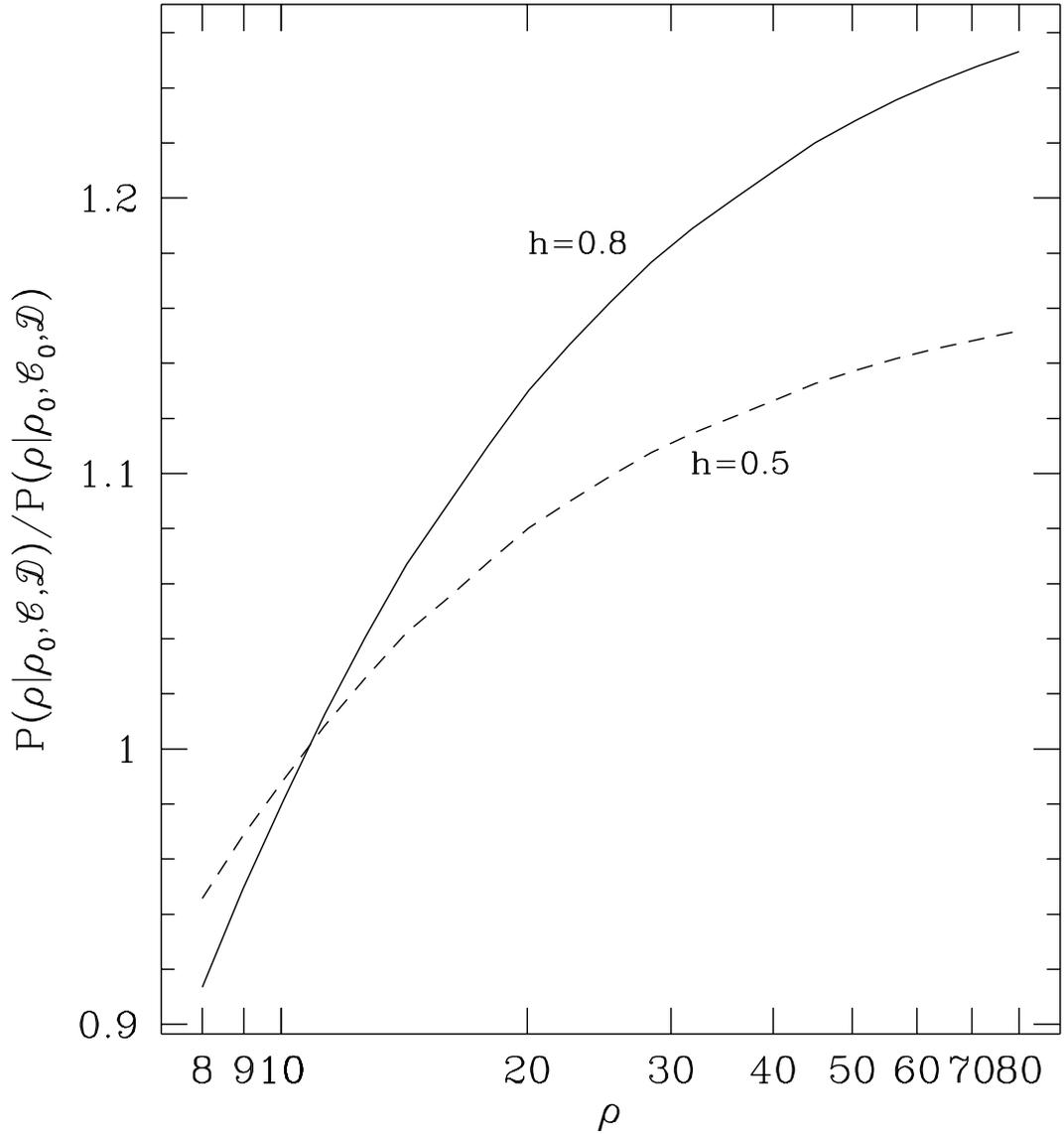}
\caption{The expected distribution of ns-ns inspiral events with
$\rho$ greater than $8$ in advanced LIGO detectors depends almost
exclusively on the Hubble parameter $h$. Shown here is the ratio of
the distribution in two matter dominated Friedmann-Robertson-Walker
cosmological models to the distribution expected in a flat and static
cosmological model. For more details see section
\protect{\ref{sec:constM0}}.}\label{fig:P(rho|rho0)}
\end{figure}

The distribution $P(\rho|\rho_0,{\cal C},{\cal D})$ is also sensitive to
$h$. Let ${\cal C}_0$ represent a flat and static cosmological
model. Then
\begin{equation}
P(\rho|\rho_0,{\cal C}_0,{\cal D}) = \left\{\begin{array}{ll}
3\rho_0^3/\rho^4&\rho>\rho_0\\
0&\rho<\rho_0
\end{array}\right.
\end{equation}
The first order correction (in $1/\rho$) owing to the expansion of the
universe depends only on the rate of expansion ($H_0$); corrections
owing to the curvature of space ($k\neq0$) enter only at second order
in $1/\rho$. Since, for ns-ns binaries and reasonable limiting
signal-to-noise $\alpha^2$ is less than $1\%$ even for advanced
interferometer designs, cosmological tests that focus only on the
observed distribution of $\rho$ will be insensitive to $q_0$ even
though the redshift to the most distant sources in the catalog is
large (this weak dependence on $q_0$ is characteristic of number-flux
cosmological tests (see, {\em e.g.,} \cite[pg. 798]{misner73a}). On
the other hand, for advanced detectors and interesting cosmological
models, the distribution $P(\rho|\rho_0,{\cal C},{\cal D})$ is
sensitive to $h$ at the 10\% level, which makes possible a measurement
of $h$ from the observations $\{\rho|\rho>\rho_0\}$ alone. Figure
\ref{fig:P(rho|rho0)} shows $P(\rho|\rho_0,{\cal
I})/P(\rho|\rho_0,{\cal C}_0,{\cal D})$ for two Einstein-deSitter
cosmological models that differ only by $h$ (in both cases ${\cal
M}_0=1.19\,\text{M}_\odot$).  The general trend is that with
increasing $h$ inspiral events are shifted toward larger $\rho$
compared to the distribution in a static universe.

\subsubsection{Examples: Uniformly distributed neutron star masses}

In a less extreme limit assume that neutron star masses are bounded
above by $m_h$ and below by $m_l$, that between these bounds their
masses are uniformly distributed, and that the component masses of an
inspiraling binary system are uncorrelated. Then ${\cal P}({\cal
M}_0)$ is given approximately by equation \ref{eqn:approx-p(m0)}, and
\widetext
\begin{mathletters}
\begin{eqnarray}
{d^2N\over dt\,d\rho}
&=& 
\left({16\over7}\pi r_0^3\dot{n}_0\right)
\left(8\over\rho\right)^4
\left(m_0\over1.2\text{M}_\odot\right)^{5/2} 
\left[\xi_{5/2}
- {100\over27}\bar{\alpha}\xi_{10/3}
+ {2\over27}\left(6q_0+85\right)\bar{\alpha}^2\xi_{25/6}\right.\nonumber\\
&&\qquad\left.{}
- {448\over33}q_0\bar{\alpha}^3\xi_{5}
+ {140\over24057}\left(1188q_0^2 + 9108q_0 -
4123\right)\bar{\alpha}^4\xi_{35/6} 
+ {\cal O}\left(\bar{\alpha}^5\right)
\right]\\
{dN\over dt}
&=& 
\left({128\over21}\pi r_0^3\dot{n}_0\right)
\left(8\over\rho_0\right)^3
\left(m_0\over1.2\text{M}_\odot\right)^{5/2} 
\left[\xi_{5/2}
- {25\over9}\bar{\alpha}_0\xi_{10/3}
+ {2\over45}\left(6q_0+85\right)\bar{\alpha}^2_0\xi_{25/6}\right.\nonumber\\
&&\qquad\left.{}
- {224\over33}q_0\bar{\alpha}^3_0\xi_{5}
+ {20\over8019}\left(1188q_0^2 + 9108q_0 - 4123\right)\bar{\alpha}^4_0\xi_{35/6}
+ {\cal O}\left(\bar{\alpha}^5_0\right)
\right]
\end{eqnarray}
where
\begin{eqnarray}
\xi_n(x_h,x_l)&=&
{2\left[
  x_h^{n+2}\left(1-x_l\right) 
+ x_l^{n+2}\left(x_h-1\right)
- \left(x_h-x_l\right)
\right]
\over
\left(n+2\right)\left(n+1\right)\left(x_h-x_l\right)
\left(x_h-1\right)\left(1-x_l\right)}
\end{eqnarray}
\narrowtext
and
\begin{eqnarray}
\bar{\alpha} &=& 
\left({8H_0r_0\over\rho}\right)
\left(m_0\over1.2\,\text{M}_\odot\right)^{5/6}\\
\bar{\alpha}_0 &=& 
\left({8H_0r_0\over\rho_0}\right)
\left(m_0\over1.2\,\text{M}_\odot\right)^{5/6}\\
x_h &=& {m_>\over m_0}\\
x_l &=& {m_<\over m_0}\\
m_< &=& m_l/2^{1/5}\\
m_0 &=& \left(m_l m_h\right)^{3/5}/\left(m_l+m_h\right)^{1/5}\\
m_> &=& m_h/2^{1/5}.
\end{eqnarray}
\end{mathletters}

\narrowtext
Over the range of $\bar{\alpha}_0$ relevant for neutron star binary
inspiral observations in LIGO-like interferometric detectors, $dN/dt$
and $d^2N/dt\,d\rho$ are only weakly dependent on $m_l$ and $m_h$ for
constant $m_0$: taking
\begin{mathletters}
\begin{equation}
x_l = 1-\epsilon
\end{equation}
we find
\begin{eqnarray}
x_h &=& 1
+ \epsilon
+ {6\over5}\epsilon^2 
+ {36\over25}\epsilon^3\nonumber\\
&&\qquad{}+ {42\over25}\epsilon^4
+ {1188\over625}\epsilon^5 + {\cal O}(\epsilon^6) \\
{dN\over dt}(\epsilon) &=& {dN\over dt}(0)
\left[
1+{21\over16}\epsilon^2 + {\cal O}\left(\alpha,\epsilon^3\right)
\right] .
\end{eqnarray}
\end{mathletters}
Observations suggest that $\epsilon$ is small \cite{finn94a}, in which
case the distribution of $\rho$ in LIGO observations is insensitive to
the finite neutron star mass range.

\subsection{Chirp mass spectrum $P({\cal M}|\rho_0,{\cal
I})$}\label{sec:spectrum} 

In a homogeneous and isotropic cosmology the rate at which binary
inspiral signals corresponding to chirp mass $\cal M$ and $\rho$
greater than $\rho_0$ are observed is
\widetext
\begin{eqnarray}
{d^2N\over dt\,d{\cal M}} &=&
\int_0^\infty dz\,
{dr\over dz}
{4\pi a_0^3 r^2\over\left(1-k r^2\right)^{1/2}}
{\dot{n}_0{\cal E}(z)\over1+z}
{{\cal P}\left({{\cal M}\over1+z}|z\right)\over1+z}
C_\Theta\left[
{\rho_0\over8}{d_L\over r_0\zeta(f_{\text{max}})}
\left(1.2\,\text{M}_\odot\over{\cal M}\right)^{5/6}
\right].
\label{eqn:dN/dtdM}
\end{eqnarray}
\narrowtext
The catalog's {\em a priori} chirp mass distribution is thus
\begin{equation}
P({\cal M}|\rho_0,{\cal C},{\cal D}) = {{d^2N/dt\,d{\cal M}}\over{dN/dt}}
\label{eqn:P(M|I)}
\end{equation}
where $dN/dt$ is given by equation \ref{eqn:dN/dt'} and ${\cal C}$
represents assumptions regarding the cosmology ($H_0$, $q_0$,
$\Omega_0$), evolution and the distribution of neutron star masses
(${\cal E}$, ${\cal P}$), and other, unenumerated, model assumptions.

As discussed in section \ref{sec:NConstraints}, observational and
theoretical evidence suggest that the ns-ns binary intrinsic chirp
mass distribution is sharply peaked. Neglecting evolution and taking
the intrinsic chirp mass of all binary systems to be a constant
${\cal M}_0$,
\widetext
\begin{eqnarray}
{d^2N\over dt\,d{\cal M}} &=& 
{\dot{n}_0{\cal E}(Z)\over {\cal M}}
{dr\over dz}(Z)
{4\pi a_0^3 r^2(Z)\over\sqrt{1-kr^2(Z)}}
C_\Theta\left[
{\rho_0\over8}{d_L(Z)\over(1+Z)^{5/6}r_0}
\left(1.2\,\text{M}_\odot\over{\cal M}_0\right)^{5/6}
\right] ,
\label{eqn:diff-explicit}
\end{eqnarray}
\narrowtext
where 
\begin{equation}
Z = {{\cal M}\over{\cal M}_0}-1.
\end{equation}

Figure \ref{fig:P(m|h,q0)} shows the distribution $P({\cal
M}|\rho_0,{\cal C},{\cal D})$ for catalogs with $\rho$ greater than 8
compiled by an advanced LIGO detector ($r_0=355\,\text{Mpc}$) in
several matter dominated FRW cosmological models. The intrinsic chirp
mass of all systems is assumed to be $1.19\,\text{M}_\odot$. Six
different models are shown, exploring two different $h$ ($0.5$ and
$0.8$) and three different $q_0$ ($1/4$, $1/2$ and $3/4$).  The
closely spaced curves with co-located extrema are of the same $h$ and
differ only in $q_0$. Note the strong dependence of $P({\cal
M}|\rho_0,{\cal C},{\cal D})$ on $h$ and the weaker, but still
significant, dependence on $q_0$: the dotted and solid curves
correspond to flat cosmological models ($q_0=1/2$), the long-dash and
dot-long-dash curves correspond to open cosmological models
($q_0=1/4$), and the short-dash and dot-short-dash curves correspond
to the closed models ($q_0=3/4$).  In general the smaller $q_0$, the
more compressed the spectrum and the smaller the tail at large ${\cal
M}$.

Since the abscissa ${\cal M}$ is related to redshift according to
${\cal M}=(1+z){\cal M}_0$ figure \ref{fig:P(m|h,q0)} also shows the
redshift of the preponderance of catalog events. For $h=0.8$ most
events are at a redshift of 9\%, while for $h=0.5$ most events are at
a redshift of 14\%.

More generally, neutron star masses are not all identical;
correspondingly, ${\cal P}({\cal M}_0)$ is not as simple as a
$\delta$-function. Modeling the neutron star mass distribution as
uniform between lower bound $m_l$ and upper bound $m_h$ leads to the
intrinsic chirp mass distribution ${\cal P}({\cal M}_0)$ given by
equation \ref{eqn:p(m0)}; the corresponding spectrum $P({\cal
M}|\rho_0,{\cal C},{\cal D})$ can be determined through equations
\ref{eqn:dN/dtdM} and \ref{eqn:P(M|I)}. Figure \ref{fig:P(m|ml,mh)}
shows the spectrum $P({\cal M}|\rho_0,{\cal C},{\cal D})$ for four
different matter-dominated Einstein-deSitter cosmological models
corresponding to two different $h$ ($0.5$ and $0.8$) and two different
neutron star mass distributions: one set of curves corresponds to the
assumption that ${\cal M}_0=1.19\,\text{M}_\odot$ for all binary
systems, while in the second set the binary component masses are
assumed to be uniformly distributed between $1.29\,\text{M}_\odot$ and
$1.45\,\text{M}_odot$. For all models shown $r_0=355\,\text{Mpc}$ and
$\rho_0=8$.

As $m_l$ and $m_h$ approach $m_0$, ${\cal P}({\cal M}_0)$ approaches
$\delta({\cal M}_0-m_0)$. The dependence of $P({\cal M}|\rho_0,{\cal
I})$ on $q_0$, shown in figure \ref{fig:P(m|h,q0)}, is similar but not
identical to its dependence on $m_h-m_l$ (for constant $m_0$):
variations in $q_0$ shift the spectrum's large $\cal M$ tail, while
increasing $m_h-m_l$ increases the spectrum's overall breadth.

\section{Conclusions}\label{sec:conclusions}

Observations of binary inspiral in a single interferometric
gravitational wave detector can be cataloged according to signal
strength (as measured by signal-to-noise ratio $\rho$) and chirp mass
$\cal M$. The distribution of events in a catalog composed of
observations with $\rho$ greater than a threshold $\rho_0$ depends on
the Hubble expansion, deceleration parameter, and cosmological
constant, as well as the distribution of component masses in binary
systems and evolutionary effects (though for neutron star binary
observations evolution is not expected to be important). In this
paper I find general expressions for the distribution with $\rho$ and
$\cal M$ of cataloged events, valid in any homogeneous and isotropic
cosmological model; I also evaluate those distributions explicitly for
matter-dominated Friedmann-Robertson-Walker models and simple models
of the neutron star mass distribution.

These distribution have two immediate, practical uses in
gravitational-wave data analysis for interferometric detectors: first,
when evaluated for with the cosmological parameters reflecting our
current best understanding of the universe, they are the prior
probabilities which, together with the matched-filtered detector
output, form the likelihood function and determine the posterior
probability that an inspiral has been detected; second, when compared
with the observed distribution in $\rho$ and $\cal M$ of many separate
binary inspiral observations, they are used to infer new and more
informed estimates for the cosmological parameters that describe the
universe.

The signal-to-noise ratio $\rho$ of a binary inspiral event is a
detector-dependent measure of the signal strength from a radiation
source that is estimated in the course of making an observation. The
normalization of $\rho$ depends on the detector's noise power spectral
density $S_n(f)$. In interferometric gravitational wave detectors like
LIGO and VIRGO the normalization involves a characteristic distance
$r_0$ and a function of the detector bandwidth $\zeta$, both of which
are depend on $S_n(f)$.

The characteristic distance $r_0$ gives an overall sense of the depth
to which the detector can ``see'' binary systems whose radiation
traverses the detector bandwidth, which is determined by
$\zeta$. Advanced LIGO interferometers are expected to be most
sensitive to binary inspiral radiation in the bandwidth
30--$200\,\text{Hz}$: over 90\% of the signal-to-noise ratio is
contributed by signal power in this narrow band. 

When searching the output of a detector for the signal from a source
an accurate model of the detector response is needed. The model need
not be accurate over the entire frequency domain, but only over that
part of the domain where the signal power overlaps with the detector
bandwidth. The dependence of $\rho$ on $\zeta$ suggests the definition
of a {\em bandwidth function} $\cal B$ for binary inspiral
observations that will be useful for determining over what range
approximate templates describing the detector response need be
accurate.

Cosmological tests based on catalogs of binary inspiral observations
with $\rho$ greater than a threshold $\rho_0$ depend on the
distribution of cataloged events with $\rho$ and $\cal M$. Tests can
make use of all the information available in the catalog or properly
constructed summaries of the cataloged events. The sensitivity of the
test depends on how the expected catalog distributions change with
changing cosmological models.

For advanced LIGO detectors, the most distant neutron star binary
inspiral events with signal-to-noise ratio greater than $8$ will arise
from distances not exceeding approximately $2\,\text{Gpc}$,
corresponding to a redshift of $0.48$ (0.26) for $h=0.8$
($0.5$). The depth is only weakly dependent on the range of neutron
star masses or the deceleration parameter. As the binary system mass
increases so does the distance it can be seen, up to a limit: in a
matter dominated Einstein-deSitter cosmological model with $h=0.8$
($0.5$) the limit is approximately $z=2.7$ (1.7) for binaries
consisting of approximately $10\,\text{M}_\odot$ black holes.

The distribution of catalog events with $\rho$ depends primarily on
$h$ and is only very weakly sensitive to either $q_0$ or the range of
neutron star masses; however, the chirp mass spectrum ({\em i.e.,} the
distribution of catalog events with $\cal M$) is very sensitive to $h$
and reasonably sensitive to both $q_0$ and the neutron star mass
range. This suggests that the spectrum is an especially powerful tool
for cosmological measurements. The dependence of the spectrum on $q_0$
and the neutron star mass distribution is similar (though not
identical), suggesting that it may be difficult to determine these
separately from observations. I am currently investigating this point
and will report on it at a later time.

\acknowledgments

I am grateful for the support of both the Alfred P.~Sloan Foundation
and the National Science Foundation (PHY~9308728).


\begin{thebibliography}{10}

\bibitem{abramovici92a}
A. Abramovici {\it et~al.}, Science {\bf 256},  325  (1992).

\bibitem{bradaschia90a}
C. Bradaschia {\it et~al.}, Nucl. Instrum. Methods Phys. Research {\bf A289},
  518  (1990).

\bibitem{thorne87a}
K.~S. Thorne,  in {\em 300 Years of Gravitation}, edited by S. Hawking and W.
  Israel (Cambridge University Press, Cambridge, 1987), pp.\ 330--458.

\bibitem{finn93a}
L.~S. Finn and D.~F. Chernoff, Phys. Rev. D {\bf 47},  2198  (1993).

\bibitem{chernoff93a}
D.~F. Chernoff and L.~S. Finn, Astrophys. J. Lett. {\bf 411},  L5  (1993).

\bibitem{schutz86a}
B.~F. Schutz, Nature (London) {\bf 323},  310  (1986).

\bibitem{markovic93a}
D. Markovi{\'c}, Phys. Rev. D {\bf 48},  4738  (1993).

\bibitem{jaranowski92a}
P. Jaranowski and A. Krolak, Astrophys. J. {\bf 394},  586  (1992).

\bibitem{cutler94a}
C. Cutler and {\'E}. Flanagan, Phys. Rev. D {\bf 49},  2658  (1994).

\bibitem{wipf95a}
S. Wipf and L.~S. Finn, Observing binary inspiral with a three-interferometer
  gravitational wave detector, in preparation.

\bibitem{narayan91a}
R. Narayan, T. Piran, and A. Shemi, Astrophys. J. Lett. {\bf 379},  L17
  (1991).

\bibitem{phinney91a}
E.~S. Phinney, Astrophys. J. {\bf 380},  L17  (1991).

\bibitem{taylor89a}
J.~H. Taylor and J.~M. Weisberg, Astrophys. J. {\bf 345},  434  (1989).

\bibitem{wolszczan91a}
A. Wolszczan, Nature (London) {\bf 350},  688  (1991).

\bibitem{prince91a}
T.~A. Prince, S.~B. Anderson, S.~R. Kulkarni, and A. Wolszczan, Astrophys. J.
  Lett. {\bf 374},  L41  (1991).

\bibitem{woosley92a}
S.~E. Woosley and T.~A. Weaver,  in {\em The Structure and Evolution of Neutron
  Stars}, edited by D. Pines, R. Tamagaki, and S. Tsuruta (Addison-Wesley,
  Redwood City, California, 1992), pp.\ 235--249.

\bibitem{finn94b}
L.~S. Finn, Phys. Rev. Lett. {\bf 73},  1878  (1994).

\bibitem{bethe95a}
H.~A. Bethe and G.~E. Brown, Astrophys. J. Lett. {\bf 445},  L129  (1995).

\bibitem{finn92a}
L.~S. Finn, Phys. Rev. D {\bf 46},  5236  (1992).

\bibitem{finn94a}
L.~S. Finn,  in {\em Proceedings of the Cornelius Lanczos International
  Centenary Conference}, {\em SIAM Proceedings Series}, edited by J.~D. Brown,
  M.~T. Chu, D.~C. Ellison, and R.~J. Plemmons (SIAM, Philadelphia, 1994), pp.\
  479--481.

\bibitem{apostolatos94a}
T.~A. Apostolatos, C. Cutler, G.~J. Sussman, and K.~S. Thorne, Phys. Rev. D
  {\bf 49},  6274  (1994).

\bibitem{cutler93b}
C. Cutler {\it et~al.}, Phys. Rev. Lett. {\bf 70},  2984  (1993).

\bibitem{kidder93a}
L.~E. Kidder, C.~M. Will, and A.~G. Wiseman, Phys. Rev. D {\bf 47},  3281
  (1993).

\bibitem{kochanek92a}
C.~S. Kochanek, Astrophys. J. {\bf 398},  234  (1992).

\bibitem{bildsten92a}
L. Bildsten and C. Cutler, Astrophys. J. {\bf 400},  175  (1992).

\bibitem{ligo89a}
R.~E. Vogt {\it et~al.}, Proposal to the {N}ational {S}cience {F}oundation,
  California Institute of Technology (unpublished).

\bibitem{saulson90a}
P.~R. Saulson, Phys. Rev. D {\bf 42},  2437  (1990).

\bibitem{gillespie94a}
A. Gillespie and F. Raab, Phys. Lett. A {\bf 190},  213  (1994).

\bibitem{gillespie94b}
A. Gillespie and F. Raab, Thermally excited vibrations of the mirrors of laser
  interferometer gravitational-wave detectors, {LIGO} preprint 94-3, 1994,
  submitted to {P}hysical {R}eview {D}.

\bibitem{weinberg72a}
S. Weinberg, {\em Gravitation and Cosmology: Principles and Applications of the
  General Theory of Relativity} (Wiley, New York, 1972).

\bibitem{misner73a}
C.~W. Misner, K.~S. Thorne, and J.~A. Wheeler, {\em Gravitation} (Freeman, San
  Francisco, 1973).

\end{thebibliography}

\begin{figure}
\epsfxsize=\columnwidth\epsffile{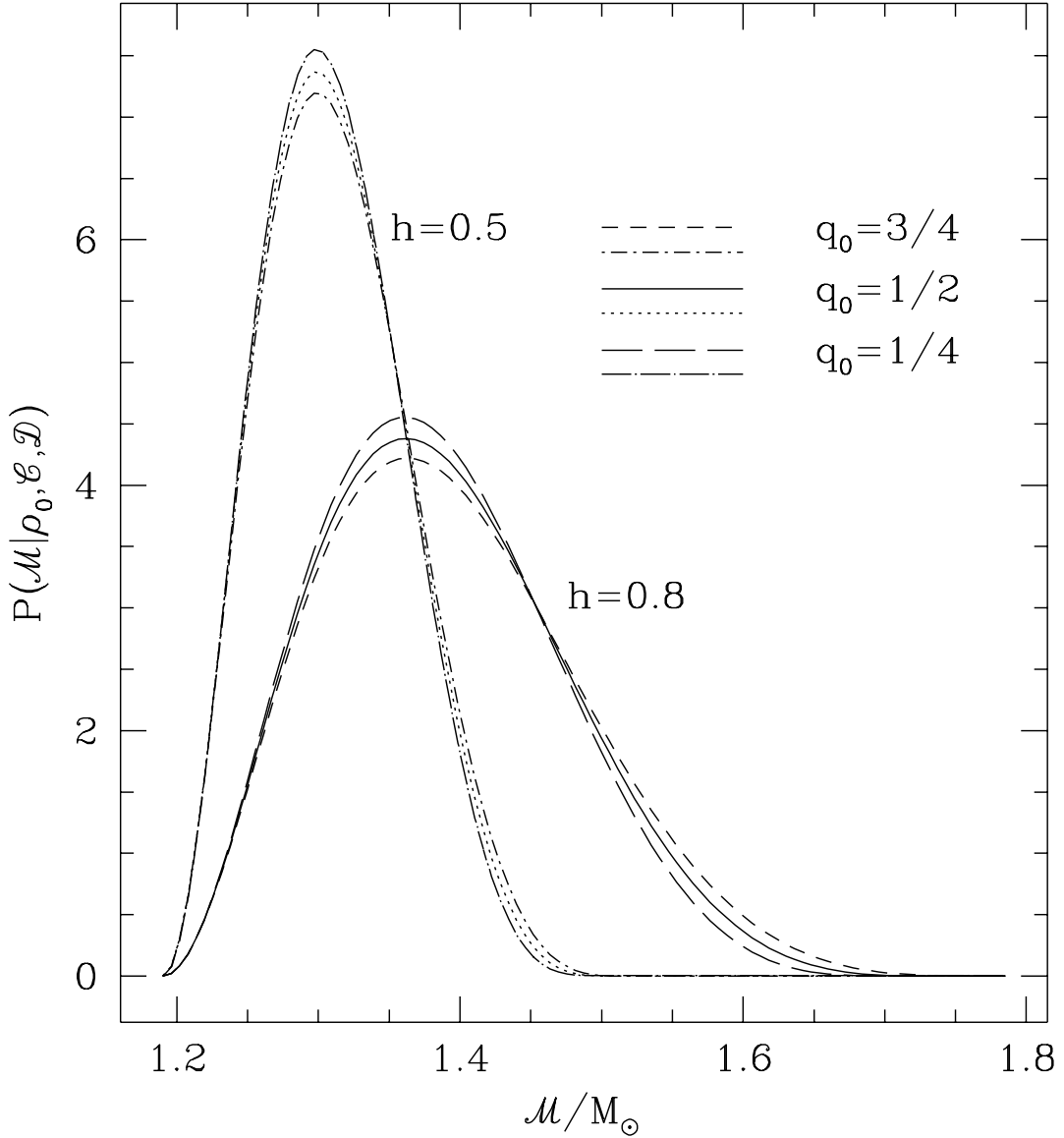}
\caption{A binary system's observed chirp mass $\cal M$ depends on its
redshift; consequently, a ns-ns binary inspiral sample will show a
range of chirp masses corresponding to the range of system
redshifts. Shown here is the expected distribution of $\cal M$ for
binary systems consisting of two $1.37\,\text{M}_\odot$ neutron stars
with $\rho>8$ in advanced LIGO detectors for open ($q_0=1/4$), flat
($q_0=1/2$) and closed ($q_0=3/4$) matter-dominated
Friedmann-Robertson-Walker cosmological models with $h=0.5$ and
$h=0.8$. In all cases, as $q_0$ increases the tail of the chirp mass
spectrum is extended. For more details, see the discussion in
\protect{\ref{sec:spectrum}}.}
\label{fig:P(m|h,q0)}
\end{figure}

\begin{figure}
\epsfxsize=\columnwidth\epsffile{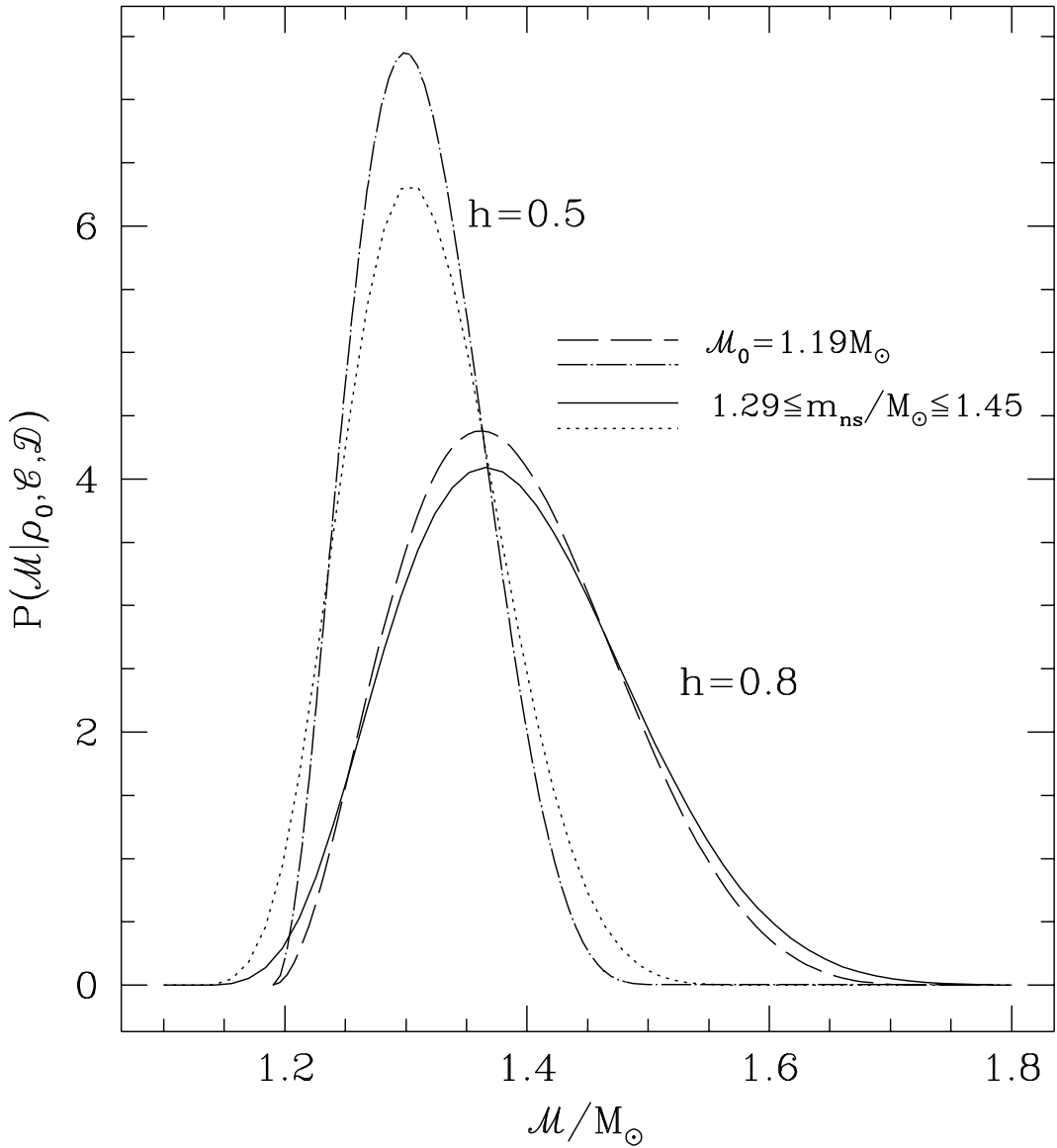}
\caption{Neutron stars do not all share the same mass, although
indications are that the mass range is small. Shown here is the
expected chirp mass distribution for observations with $\rho>8$ made
in an advanced LIGO detector for two different matter-dominated
Einstein-deSitter cosmological models ($q_0=1/2$ and $h=0.5$ or
$h=0.8$) and two different neutron star mass distributions. In the
first distribution all binaries are assumed to have intrinsic chirp
mass ${\cal M}_0=1.19\,\text{M}_\odot$, while in the second the binary
component masses are assumed to be uniformly distributed between lower
bound $1.29\,\text{M}_\odot$ and upper bound
$1.45\,\text{M}_\odot$. As the mass distribution broadens, the chirp
mass spectrum also broadens. It does so nearly symmetrically; in
contrast, variations in $q_0$ for fixed mass distribution (shown in
figure \protect{\ref{fig:P(m|h,q0)}}) alter the large $\cal M$ tail of
the spectrum, leaving the small ${\cal M}$ tail essentially unchanged.
For more details see section
\protect{\ref{sec:spectrum}}.}\label{fig:P(m|ml,mh)}
\end{figure}

\end{document}